\documentclass[aps,a4paper,twocolumn]{revtex4}

\usepackage{amsmath}
\usepackage{graphicx}
\usepackage{subfigure}
\usepackage[usenames,dvipsnames]{color}
\definecolor{darkblue}{RGB}{0,0,196}
\usepackage[colorlinks=true,linkcolor=darkblue,citecolor=darkblue,urlcolor=darkblue]{hyperref}
\usepackage{setspace}
\usepackage{hyperref}
\usepackage{xcolor}
\usepackage{enumerate}% http://ctan.org/pkg/enumerate
\hypersetup{
  colorlinks   = true, %Colours links instead of ugly boxes
  urlcolor     = blue, %Colour for external hyperlinks
  linkcolor    = blue, %Colour of internal links
  citecolor   = blue %Colour of citations
}
\usepackage{graphicx}
\usepackage{amsmath,bbm}
\usepackage{amssymb,bm}
\usepackage{yfonts}
\usepackage{comment}

\begin{document}

\title{Effects of clustered nuclear geometry on the anisotropic flow in O-O collisions at the LHC within a multiphase transport model framework}

\author{Debadatta Behera}

\author{Suraj Prasad}

\author{Neelkamal Mallick}

\author{Raghunath Sahoo\footnote{Corresponding author}}
\email{raghunath.sahoo@cern.ch}

\affiliation{Department of Physics, Indian Institute of Technology Indore, Simrol, Indore 453552, India}

\begin{abstract}
To understand the true origin of flowlike signatures and applicability of hydrodynamics in small collision systems, effects of soft QCD dynamics, the sensitivity of jetlike correlations, and nonequilibrium effects, efforts are being made to perform \textit{p}-O and O-O collisions at the LHC and RHIC energies. It is equally interesting to look into the possible signatures of an $\alpha$-clustered nuclear geometry in $^{16}$O-$^{16}$O collisions by studying the initial-state effects on the final-state observables. In this work, within a multiphase transport model, we implement an $\alpha$-cluster tetrahedral density profile in the oxygen nucleus along with the default Woods-Saxon density profile. We study the eccentricity ($\epsilon_2$), triangularity ($\epsilon_3$), normalized symmetric cumulants [NSC(2,3)], elliptic flow ($v_2$), and triangular flow ($v_3$) in $^{16}$O-$^{16}$O collisions at $\sqrt{s_{\rm NN}} = 7~$TeV. The constituent quark number scaling of the elliptic flow is also reported. For the most central collisions, enhanced effects in $\langle \epsilon_3 \rangle/ \langle \epsilon_2 \rangle$ and $\langle v_3 \rangle/ \langle v_2 \rangle$ with a negative value of NSC(2,3), and an away-side broadening in the two-particle azimuthal correlation function [$C(\Delta \phi)$] of the identified particles are observed in the presence of an $\alpha$-clustered geometry. 
\end{abstract}
\date{\today}
\maketitle 

\section{Introduction} 
\label{intro}
Ultrarelativistic heavy-ion collisions at the Large Hadron Collider (LHC) and the Relativistic Heavy Ion Collider (RHIC) create high temperature and density, which provide suitable conditions for producing a locally thermalized and deconfined partonic medium. This hot and dense fireball is made up of QCD matter, i.e., quarks and gluons,  and thus called quark-gluon plasma (QGP). Studies related to QGP investigate all the indirect signatures as QGP is a highly short-lived state due to the behavior of strongly interacting matter. QGP expands rapidly, and its evolution is well understood through relativistic viscous hydrodynamics with dissipative effects. Thus, the initial-state collision geometry and the fluctuations in energy and entropy density are embedded in the final-state multiparticle correlations through this collective expansion of the QGP~\cite{Heinz:2013th, Voloshin:2008dg, Ollitrault:1992bk}. Usually, this is studied as the medium response to the initial eccentricity ($\epsilon_2$) and triangularity ($\epsilon_3$) by quantifying the Fourier coefficients ($v_2$ and $v_3$) of the azimuthal momentum distribution of the final-state hadrons~\cite{voloshin-zhang}. Experimental measurements of these flow coefficients agree with the predictions from hydrodynamic calculations suggesting that QGP behaves like a perfect fluid~\cite{STAR:2005gfr}. Thus, the presence of finite-flow coefficients is considered a signature of the hydrodynamic behavior of the QGP and hence, the thermalization in the early stages of the collision. Recently, similar signatures have been observed in small collision systems such as the high-multiplicity $pp$ collisions, where hydrodynamic expansion or collectivity is usually not expected~\cite{CMS:2016fnw}. These observations also raise questions on the applicability of hydrodynamics in small collision systems formed in ultrarelativistic nuclear collisions. As the system size of $^{16}$O-$^{16}$O overlaps high-multiplicity $pp$ and peripheral Pb-Pb collisions, it provides an opportunity to explore the origin of flowlike signatures in small collision systems. \\

Another interesting direction is to explore how the final-state observables are affected by the initial-state nuclear structure, nuclear shape deformation, or even the presence of $^4$He-nuclei (known as $\alpha$-clusters) inside the nucleus of elements having $4n$-number of nucleons, such as $^8$Be, $^{12}$C, and $^{16}$O, to name a few. Studies related to nuclear shape deformation have been carried out at the RHIC~\cite{Giacalone:2021udy, Haque:2019vgi, PHENIX:2018lia} and at the LHC with Xe-Xe collisions at $\sqrt{s_{\rm  NN}} = 5.44$~TeV~\cite{ALICE:2018lao, CMS:2019cyz, ATLAS:2019dct}. Results show quadruple deformation in $^{129}$Xe nucleus. The presence of four $^4$He--clusters inside the $^{16}$O nucleus was first proposed by Gamow back in 1930s~\cite{gamow} and then by Wheeler~\cite{Wheeler:1937zza}. Although there is evidence for the existence of such a clustered structure~\cite{Bijker:2014tka, Wang:2019dpl, He:2014iqa, He:2021uko}, the contribution of the clustered states in the ground state of $^{16}$O was found to be less than 30\%~\cite{Zuker:1968zz}. Recently, there are proposals for dedicated runs for $^{16}$O-$^{16}$O collisions at both RHIC and LHC~\cite{Brewer:2021kiv,Katz:2019qwv}. This could clarify the origin of collectivity on small systems and the effects of clustered nuclear geometry on the final-state observables. \\

In recent years, there have been several theoretical studies reported on oxygen collisions based on Glauber Monte Carlo~\cite{Rybczynski:2019adt, Sievert:2019zjr, Huang:2019tgz}, different hydrodynamic models~\cite{Lim:2018huo, Summerfield:2021oex, Schenke:2020mbo}, global observables~\cite{Behera:2021zhi}, parton energy loss~\cite{Huss:2020whe}, and jet-quenching effects across small to large collision systems~\cite{Zakharov:2021uza}. Some observables showing evidence of the signatures of $\alpha$-clusters are also reported in $^{16}$O-$^{16}$O collisions~\cite{Li:2020vrg,Ding:2023ibq, Wang:2021ghq, Rybczynski:2017nrx,alpha-frag}. In this work, within a multiphase transport model, we implement an $\alpha$-cluster tetrahedral density profile in the oxygen nucleus along with the default Woods-Saxon density profile. We study the eccentricity ($\epsilon_2$), triangularity ($\epsilon_3$), normalized symmetric cumulants [NSC(2,3)], elliptic flow ($v_2$), and triangular flow ($v_3$) in $^{16}$O-$^{16}$O collisions at $\sqrt{s_{\rm NN}} = 7~$TeV.\\

In addition, the elliptic flow coefficients as a function of transverse momentum ($v_2(p_{\rm T})$) for the light-flavor hadrons such as $\pi^{\pm}$, $K^{\pm}$, and $p+\bar{p}$ are studied for nuclear collisions with default Woods-Saxon and $\alpha$-cluster tetrahedral density profiles. The appearance of hadronic collectivity is believed to have originated from the early deconfined partonic phase and is subsequently transferred to the hadrons via the quark recombination mechanism of hadronization. This is also known as the quark coalescence model~\cite{Voloshin:2002wa}. This behavior leads to the observation of a higher flow of baryons than mesons in the intermediate $p_{\rm T}$, and ideally to the number-of-constituent-quark (NCQ) scaling~\cite{Molnar:2003ff, Sato:1981ez, Dover:1991zn}. Experimentally, at RHIC, the NCQ scaling seems to be valid as seen in Au-Au collisions at $\sqrt{s_{\rm NN}} = 200$~GeV~\cite{STAR:2003wqp, PHENIX:2003qra}. However, at the LHC energies, the scaling is only approximate~\cite{ALICE:2010suc, ALICE:2014wao, ALICE:2022zks}. Using AMPT in string melting mode, the NCQ scaling is observed at the top RHIC energies in Au-Au collisions~\cite{Singha:2016aim}. However, NCQ scaling seems to be violated using the same model at the LHC energies in Pb-Pb. The breaking of NCQ scaling in AMPT string melting mode is found to be independent of the magnitude of parton-parton cross sections and hadron cascade time~\cite{Singha:2016aim}. However, the breaking of scaling is understood as the increase in the partonic density at the LHC energy in Pb-Pb collisions. Further, Si-Si collisions at this energy show NCQ scaling, which adds to this understanding~\cite{Singha:2016aim}. This makes the case appealing to look for the validation of NCQ scaling in $^{16}$O-$^{16}$O collisions at $\sqrt{s_{\rm NN}} = 7~$TeV. In a recent event-shape dependent study of NCQ scaling using transverse spherocity ($S_0$) in heavy-ion collisions in AMPT, it is reported that low-$S_0$ (jetty-like) events show more deviation from the NCQ scaling than the $S_0$-integrated (unbiased) events~\cite{Mallick:2021hcs}. In Pb-Pb collisions, the deviation appears in the $S_0$-integrated events and gets enhanced in low-$S_0$ events, whereas in Au-Au collisions, the scaling violation appears only in the low-$S_0$ events and not in the $S_0$-integrated events. These results show the dependence of NCQ scaling on the event shapes, and it awaits experimental confirmation. For the time being, we proceed to study the NCQ scaling behavior in $^{16}$O-$^{16}$O collisions at $\sqrt{s_{\rm NN}} = 7~$TeV using the AMPT string melting model, and explore the possible role of density profiles on the NCQ scaling in small collision systems. Further, for the most central collisions, we observe enhanced effects in $\langle \epsilon_3 \rangle/ \langle \epsilon_2 \rangle$ and $\langle v_3 \rangle/ \langle v_2 \rangle$ with a negative value of NSC(2,3), and an away-side broadening in the two-particle azimuthal correlation function [$C(\Delta \phi)$] of the identified particles. Here onwards, for the sake of simplicity, we write O-O instead of $^{16}$O-$^{16}$O throughout the text.\\

The paper is organized as follows. It begins with a brief introduction to the event generator(a multiphase transport model), the $\alpha$-cluster geometry implementation, and estimation of anisotropic flow coefficients via the two-particle correlation method in Sec.~\ref{section2}. The paper then shows and describes the results for eccentricity, triangularity, elliptic flow, triangular flow, and the number-of-constituent-quarks scaling of the elliptic flow in Sec.~\ref{section3}. Finally, the paper concludes with the important findings summarized in Sec.~\ref{section4}.

\section{Event Generation and Analysis Methodology}
\label{section2}
In this section, we briefly introduce the components of the AMPT model, the tuning used to generate the collisions, and the implementation of the $\alpha$-cluster geometry in the oxygen nucleus. The two-particle correlation method used to estimate the flow coefficients is also discussed.

\subsection{A multiphase transport model}
 A multiphase transport model (AMPT) is a Monte Carlo-based transport model for heavy-ion collision. It consists of four main stages: initialization of the collisions, parton cascade, hadronization, and hadron transport \cite{Zhang:1999mqa,Zhang:2000bd,Zhang:2000nc,Lin:2001cx,Lin:2001yd,Pal:2001zw,Lin:2001zk,Zhang:2002ug,Pal:2002aw,Lin:2002gc,Lin:2003ah,Lin:2003iq,Wang:1991hta,Zhang:1997ej,He:2017tla,Li:2001xh,Greco:2003mm}. The initialisation of the collisions is performed by HIJING, where the differential cross-section of produced minijets in pp collisions is converted into AA and p-A collisions~\cite{Wang:1991hta}. The parton cascade or the parton transport is performed by Zhang's Parton Cascade (ZPC) model~\cite{Zhang:1997ej}. In the string-melting version of AMPT, the coloured strings are transformed into the low momentum partons. The transported partons are hadronized using a spatial coalescence mechanism in the string-melting version of AMPT \cite{Lin:2001zk,He:2017tla}; however, in the default version of AMPT, the Lund string fragmentation mechanism is used to perform the hadronisation. A relativistic transport model is used for the evolution of the produced hadrons~\cite{Li:2001xh,Greco:2003mm}. In the current work, we have used the string melting mode of AMPT (version 2.26t9b) since the quark coalescence mechanism well describes the particle flow and spectra at the mid-transverse momentum region~\cite{Fries:2003vb,Fries:2003kq}.  The choice of centrality selection is taken from the geometrical slicing based on the impact parameter distribution for both the Woods-Saxon and $\alpha$-cluster density profile of the nuclei.  The impact parameter cuts for different centralities are the same as in Ref.~\cite{Behera:2021zhi}, which are obtained via the Glauber model estimation. The AMPT settings for the O-O system in the current work are also the same as reported in Ref.~\cite{Behera:2021zhi}.\\

 In heavy-ion collisions, the typical density profile for a nucleus is considered to be the Woods-Saxon distribution. The Wood-Saxon charge density is given in terms of a three-parameter fermi (3pF) distribution as,

 \begin{equation}
\rho (r) = \frac{\rho_{0}(1+w(\frac{r}{r_0})^2)}{1+{ \rm exp}(\frac{r-r_0}{a})}.
\end{equation}

Here, $r$ is the radial distance from the center of the nucleus, $a$ is the skin depth of the nucleus, $r_{0}$ is the mean radius of the nucleus, and $w$ is the deformation parameter. In the oxygen nucleus, $r_{0}$ = 2.608 $\rm{fm}$, $a$ is 0.513 $\rm{fm}$ and  $w$ is -0.051~\cite{Loizides:2014vua}. $\rho_{0}$ is the nuclear density constant at $r = 0$.\\

%Since several studies have predicted the signal of $\alpha$-cluster structure inside O$^{16}$ nucleus, these studies have motivated us to implement $\alpha$-cluster structure inside the oxygen nucleus using the AMPT model.

We also implement the $\alpha$-cluster structure inside the oxygen nucleus using the AMPT model. The implementation is done numerically by creating a geometric distribution of a regular tetrahedral structure having $^4$He nuclei placed at the vertices.
For the $^4$He nucleus, the distribution of nucleons follows the Wood-Saxon density profile described in Eq. 1 with the parameters $r_0$ = 0.964 $\rm{fm}$, $a$ = 0.322 $\rm{fm}$, and $w$ = 0.517. This leads to the $\rm{rms}$ radius for $^4$He nucleus to be 1.676 $\rm fm$. These $\alpha$-clustered nuclei are positioned on the vertices of a standard tetrahedron with a side length of 3.42 $\rm{fm}$. In this configuration, the $\rm{rms}$ radius for $^{16}$O is calculated to be 2.699 $\rm{fm}$~\cite{Behera:2021zhi,Li:2020vrg}. The orientation of the tetrahedron is randomized for each projectile and target on an event-by-event basis.

\subsection{Two-particle correlation method}

In noncentral heavy-ion collisions, the collision overlap region is anisotropic in space. The pressure gradient of the thermalized partonic medium formed in such collisions can transform the initial spatial anisotropies into the momentum space azimuthal anisotropies. These azimuthal anisotropies of different orders can be quantified by the coefficients of Fourier series decomposition of the momentum distribution of final-state particles, given as:

\begin{eqnarray}
E\frac{d^3N}{dp^3}=\frac{d^2N}{2\pi p_{\rm T}dp_{\rm T}dy}\bigg(1+2\sum_{n=1}^\infty v_n \cos[n(\phi -\psi_n)]\bigg)\,.\nonumber\\
\label{eq2}
\end{eqnarray}

Here, $\phi$ represents the azimuthal angle of the final-state particles in the transverse plane, and $\psi_{n}$ represents the $n$th harmonic event plane angle~\cite{v2eventplane}. $v_n$ is the $n$th-order anisotropic flow coefficient where $n$ = 1 stands for directed flow, $n$ = 2 is the elliptic flow and $n$ = 3 quantifies the triangular flow. Anisotropic flow coefficients of different orders can be estimated as follows:

 \begin{eqnarray}
v_{n} = \langle \cos[n(\phi - \psi_n)]\rangle
\label{eq3}
\end{eqnarray}

In experiments, obtaining the event plane angle is not trivial, and Eq.~\ref{eq3} includes the non-flow effects, such as contributions from resonance decays and jets. On the other hand, a two-particle correlation method to estimate the flow coefficients can, not only efficiently reduce the non-flow contribution by implementing a proper pseudorapidity gap but also does not require the event-plane angle. In this study, we have ignored the pseudorapidity dependence of $\psi_n$, which is observed in the experiments.\\

 To estimate the anisotropic flow coefficients using the two-particle correlation method, one requires the two-particle correlation function, which can be determined using the following steps \cite{ATLAS:2012at}:
 
\begin{enumerate}[(i)]
    \item In each event, two sets of particles are formed based on their transverse momenta, namely, ``a" and ``b". ``a" denotes the trigger particles, whereas ``b" represents the associated particle set.
    \item Each particle from trigger group (``a") pairs with each particle from associate group (``b") and the relative pseudorapidities ($\Delta\eta=\eta_a-\eta_b$) and relative azimuthal angles ($\Delta\phi=\phi_a-\phi_b$) are determined.
    \item Same event pairs ($S(\Delta\eta,\Delta\phi)$) and mixed event pairs are ($B(\Delta\eta,\Delta\phi)$) are determined. In the same event pair, both ``a" and ``b" belong to the same event; however, in the mixed event pair, ``a" and ``b" are from different events where ``a" pairs with ``b" from five randomly selected events to remove physical correlations.
    \item Two particle correlation function ($C(\Delta\eta,\Delta\phi)$) is determined by taking the ratio of $S(\Delta\eta,\Delta\phi)$ to $B(\Delta\eta,\Delta\phi)$.
\end{enumerate}

 In this analysis, we use final-state charged hadrons with kinematic cuts as $|\eta|<$ 2.5 and $p_{\rm T}>$ 0.4 GeV/$c$ to encompass a broader spectrum of particles. To omit the jet peak region seen in the $C(\Delta\eta,\Delta\phi)$ distribution, the $\Delta\eta$ interval is carefully chosen. The interval, in our case, is implemented to be $1.0 <|\Delta \eta|< 4.8$ to obtain 1D correlation $C(\Delta\phi)$, given as:

\begin{eqnarray}
C(\Delta\phi)= \frac{dN_{\rm pairs}}{d\Delta\phi} = A \times \frac{\int S(\Delta\eta ,\Delta\phi) d\Delta\eta}{\int B(\Delta\eta ,\Delta\phi)d\Delta\eta}.
\label{eq5}
\end{eqnarray}

Here, the normalization constant $A$ ensures that at a given $\Delta\eta$ interval, there is the same number of pairs in the same events and mixed events.

\begin{figure*}[ht!]
\includegraphics[scale=0.28]{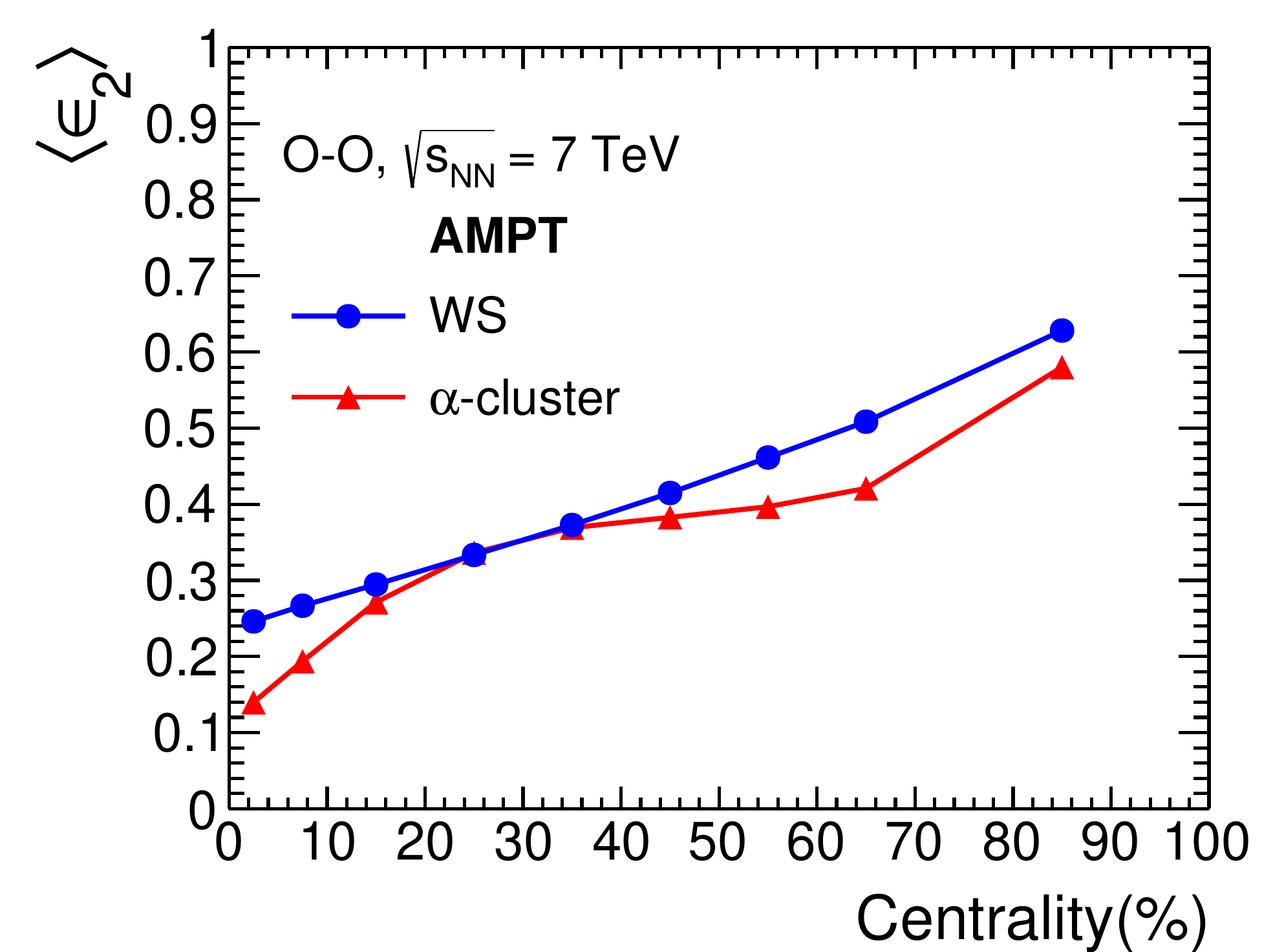}
\includegraphics[scale=0.28]{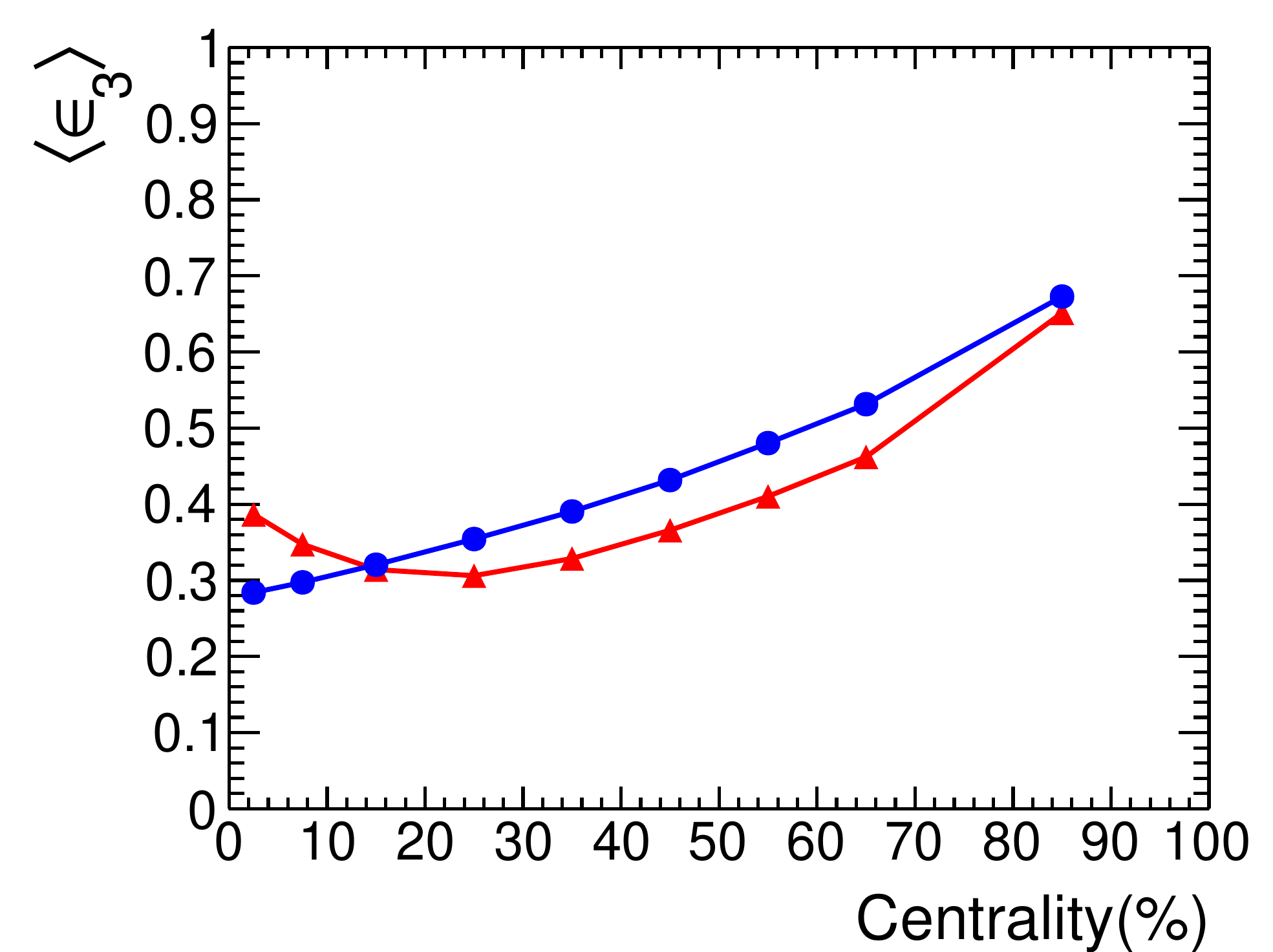}
\includegraphics[scale=0.28]{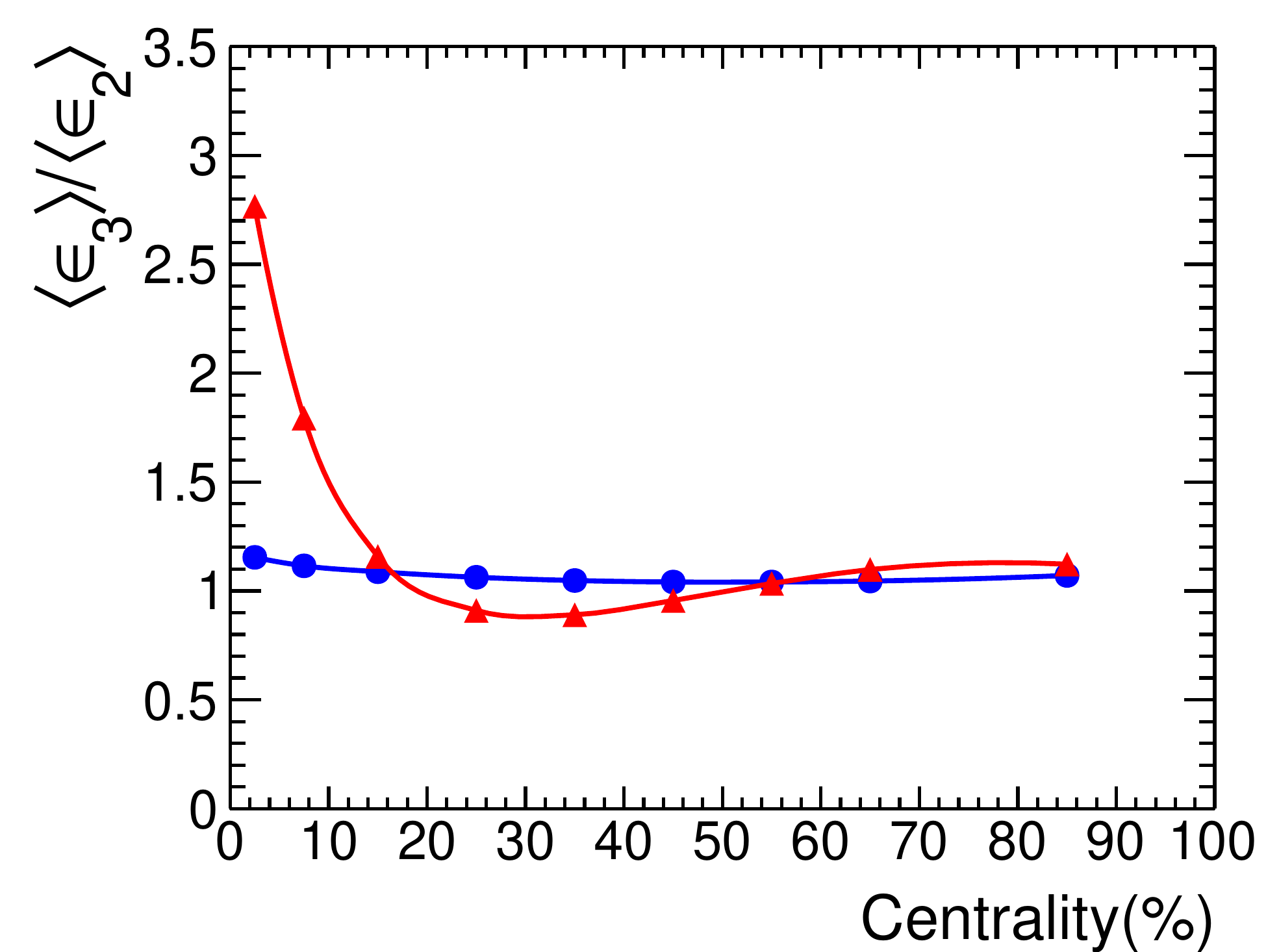}
\caption[]{(Color online) Centrality dependence of average eccentricity ($\langle\epsilon_{2}\rangle$), triangularity ($\langle\epsilon_{3}\rangle$), and $\langle\epsilon_{3}\rangle / \langle\epsilon_{2}\rangle$ for O-O collisions at $\sqrt{s_{\rm NN}} = 7$~TeV in the AMPT string melting model for both Woods-Saxon and $\alpha$-cluster type nuclear density profiles.}
\label{ecc}
\end{figure*}

The pair distribution ($N_{\rm pairs}$) or 1D correlation function can be expanded into a Fourier transform in $\Delta\phi$ as follows:

\begin{eqnarray}
C(\Delta\phi) = \frac{dN_{\rm pairs}}{d\Delta\phi} \propto  \bigg[1+2\sum_{n=1}^\infty v_{n,n}(p_{\rm T}^{a},p_{\rm T}^{b}) \cos (n\Delta\phi)  \bigg].
\label{eq6}
\end{eqnarray}

Here, $v_{n,n}$ is the two-particle flow coefficient. In this definition, the convolution of particle pairs removes the event plane angle. Now, $v_{n,n}$ can be obtained as:

\begin{eqnarray}
v_{n,n}(p_{\rm T}^{a}, p_{\rm T}^{b}) = \langle cos(n\Delta\phi) \rangle
\label{eq8}
\end{eqnarray}

 In terms of $p_{\rm T}^{a}$ and $p_{\rm T}^{b}$, $v_{n,n}$ are symmetric functions. The definition of harmonics in Eq.~\ref{eq2} enters to Eq.~\ref{eq6}, which can be written as:

\begin{eqnarray}
\frac{dN_{\rm pairs}}{d\Delta\phi} \propto  \bigg[ 1+2\sum_{n=1}^\infty v_{n}(p_{\rm T}^{a}) v_{n}(p_{\rm T}^{b}) \cos (n\Delta\phi)  \bigg].
\label{eq9}
\end{eqnarray}

If collective expansion is what causes azimuthal anisotropy, then $v_{n,n}$ can be factorized into the product of two single-particle harmonic coefficients. 

\begin{eqnarray}
v_{n,n}(p_{\rm T}^{a},p_{\rm T}^{b})= v_{n}(p_{\rm T}^{a}) v_{n}(p_{\rm T}^{b}).
\label{eq10}
\end{eqnarray}

From Eq.~\ref{eq10}, $v_{n}$ can be estimated as:
\begin{eqnarray}
v_{n}(p_{\rm T}^{a})= v_{n,n}(p_{\rm T}^{a},p_{\rm T}^{b})/\sqrt{v_{n,n}(p_{\rm T}^{b},p_{\rm T}^{b})}
\label{eq11}
\end{eqnarray}

Following the above steps, one can obtain the nth-order coefficients of the azimuthal anisotropy of all-charged particles along with identified particles such as $\pi^{\pm}$, $K^{\pm}$, and $p+\bar{p}$ for the O-O collision system at the LHC energies using the AMPT model.

\begin{figure}[ht!]
\includegraphics[scale=0.3]{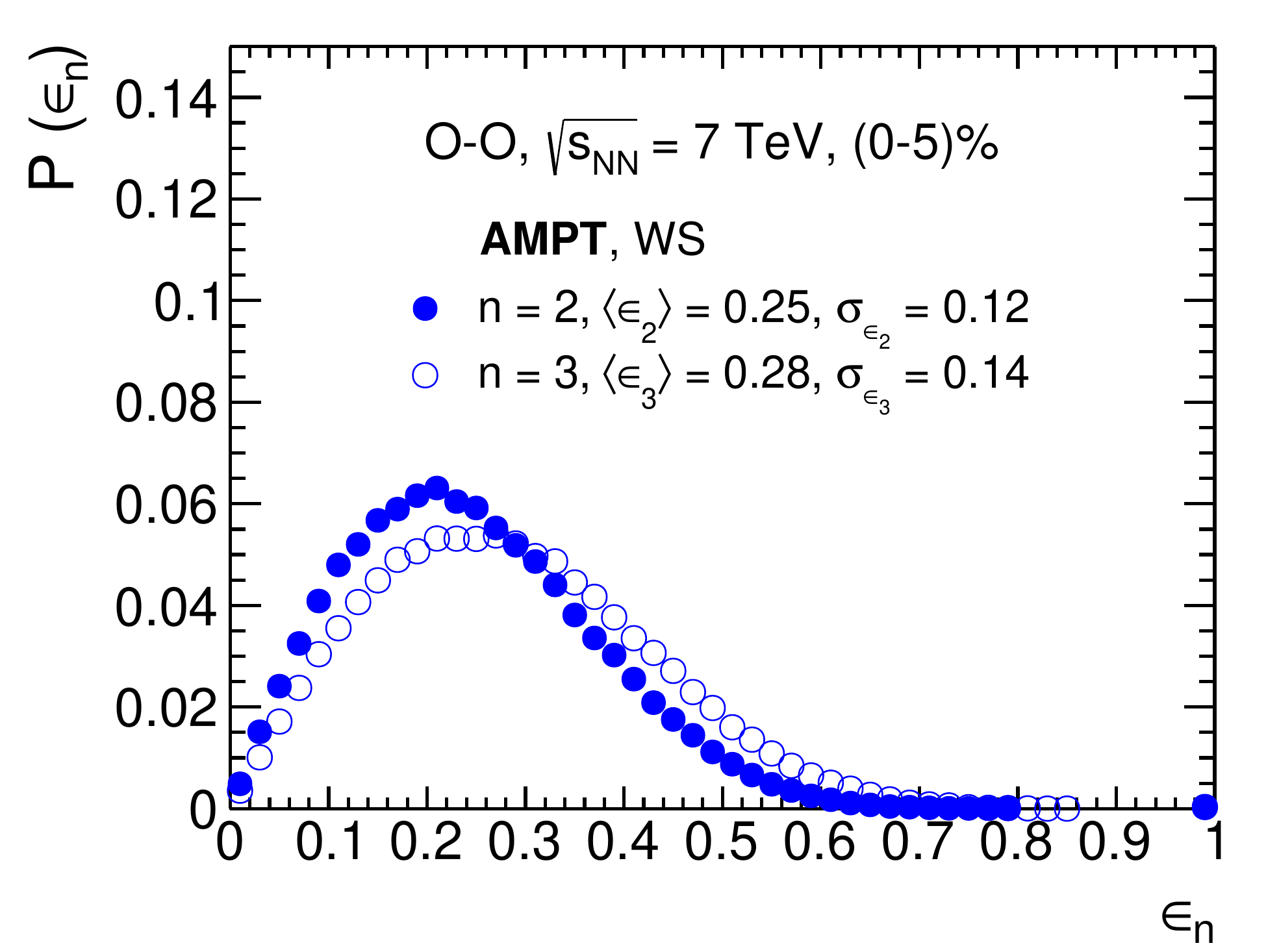}
\includegraphics[scale=0.3]{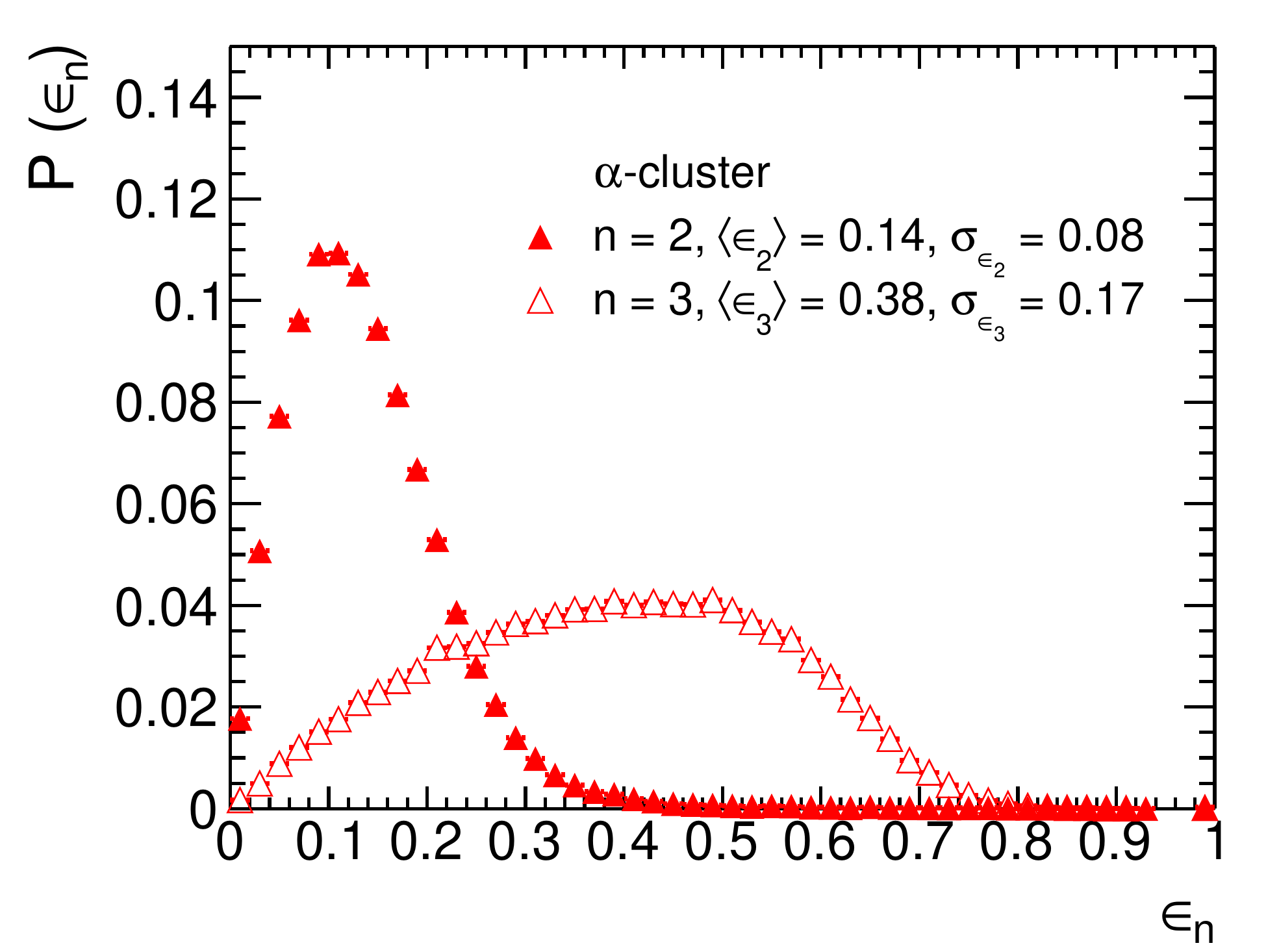}
\caption[]{(Color online) Eccentricity ($ \langle\epsilon_{2}\rangle $) and triangularity ($ \langle\epsilon_{3}\rangle $) distribution for the most central case in O-O collisions at $\sqrt{s_{\rm NN}}$ = 7 TeV in Woods-Saxon (top) and $\alpha$-cluster (bottom) type nuclear density profiles.}
\label{eccdist}
\end{figure}

\section{Results and Discussions}
\label{section3}
In this section, we start by discussing the results of the participant eccentricity, triangularity, and the correlations among them using normalized symmetric cumulants for both the Woods-Saxon density profile and the $\alpha$-clustered structure. Then we discuss the evolution of elliptic and triangular flow with centrality, their ratios, and their scalings with the initial eccentricities of the same order. We discuss the two-particle azimuthal correlation function for the identified hadrons. Finally, the elliptic flow as a function of transverse momentum and their NCQ scaling with transverse kinetic energy is discussed for different centralities and nuclear profiles.

\subsection{Eccentricity and triangularity}

In a collision of two nuclei, the overlap region of the colliding nucleons is not spherical and isotropic. It majorly depends upon the colliding nuclei species, the centrality of the collision, and the distribution of the nucleons inside the nucleus. Eccentricity represents the elliptic shape of the overlap region of the colliding nucleons and is purely geometric; however, triangularity represents the triangular shape of the region, and it arises due to event-by-event density fluctuations in the collision overlap region \cite{Prasad:2022zbr}. The anisotropic flow coefficients of the final-state hadrons have a significant contribution from the initial geometrical anisotropies. Eccentricity greatly influences the elliptic flow; however, the influence of triangularity on triangular flow is limited only to (65-70)\% for a minimally viscous fluid \cite{Chaudhuri:2011pa}. The study of eccentricity, triangularity, elliptic flow, and triangular flow may unveil information about the medium response to different harmonic flow coefficients. It is not trivial to determine the eccentricity and triangularity in experiments; however, it can be estimated in the AMPT model using the following expression \cite{Prasad:2022zbr, Petersen:2010cw}:

\begin{equation}
\epsilon_{\rm n}=\frac{\sqrt{\langle{r^{\rm n}\cos(n\phi_{\text{part}})}\rangle^{2}+\langle{r^{\rm n}\sin(n\phi_{\text{part}})}\rangle^{2}}}{\langle{r^{\rm n}}\rangle}
\label{eq:eccentricity}
\end{equation}
where $r$ and $\rm{\phi_{\text{part}}}$ are the polar co-ordinates of the participants. In $\epsilon_{n}$, $\rm n$ = 2 corresponds to eccentricity ($\epsilon_{2}$) and $\rm n$ = 3 corresponds to triangularity ($\epsilon_{3}$). In Fig.~\ref{ecc}, the event averaged eccentricity ($\langle\epsilon_2\rangle$) (left), triangularity ($\langle\epsilon_3\rangle$) (middle), their ratios (right) determined from AMPT for the Woods-Saxon density profile and $\alpha$-clustered structure in O-O collisions at $\sqrt{s_{\rm NN}}$ = 7 TeV are shown. As traditionally observed in heavy-ion collisions, both initial nucleon distributions have similar behavior of $\langle\epsilon_2\rangle$ with centrality. The value of $\langle\epsilon_2\rangle$ is observed to be increasing towards the peripheral collisions as the overlap region gets largely elliptic with increasing the impact parameter of the collisions. However, for a given centrality class, $\langle\epsilon_2\rangle$ is lower for $\alpha$-cluster case compared to Woods-Saxon nuclear density profile except for the mid-central cases where both of the profiles have similar values of $\langle\epsilon_2\rangle$. This indicates that even if the number of participants in a collision is similar, the distribution of the nucleons inside the nucleus significantly contributes to the eccentricity, which is expected to finally be reflected in the anisotropic flow coefficients given the hydrodynamical behavior of the medium formed.
A similar trend of $\langle\epsilon_3\rangle$ is observed as a function of centrality where the mean triangularity for both Woods-Saxon and $\alpha$-cluster density profiles is increasing towards the peripheral collisions owing to the appearance of a more triangular shape. This trend of $\langle\epsilon_3\rangle$ as a function of centrality has a peculiar behavior for the   $\alpha$-cluster case where the value decreases from central to mid-central collisions, attains a minimum and then starts to rise again towards the peripheral collisions. Nevertheless, the value for the Woods-Saxon nuclear density profile dominates over the $\alpha$-cluster structure throughout the centrality selection except for the most central cases, i.e. (0--5)\% and (5--10)\%. The $\alpha$-cluster structure thus can have more significant event-by-event fluctuations in the participant distribution due to its larger triangularity in the most central collisions. It is to be noted that, due to a smaller collision system, the number of sources that contribute to $\epsilon_n$ decreases, which can make $\epsilon_n$ more significant in O-O collisions compared to Pb-Pb or Au-Au collision systems \cite{Bzdak:2013rya}. A similar study is reported in Ref.~\cite{Li:2020vrg} for the most central ($b = 0$) O-O collisions at $\sqrt{s_{\rm NN}}$ = 6.37 TeV using AMPT, where the reported values for $\langle\epsilon_2\rangle$ and $\langle\epsilon_3\rangle$ follow a similar trend for a given choice of the nuclear profile. However, the values of $\langle\epsilon_2\rangle$ in Ref.~\cite{Li:2020vrg} are larger as compared to the present
study, whereas $\langle\epsilon_3\rangle$ values are almost comparable. This could be because of the use of the initial partons to estimate $\langle\epsilon_2\rangle$ and $\langle\epsilon_3\rangle$, in contrast to the present study, where we use the initial participant nucleons.\\

The right-most panel of Fig.~\ref{ecc} shows the ratio of mean triangularity to the eccentricity, i.e., $\langle\epsilon_3\rangle/\langle\epsilon_2\rangle$ as a function of centrality. Here the ratio is plotted for both Woods-Saxon nuclear density profile and $\alpha$-clustered structure in O-O collisions at $\sqrt{s_{\rm NN}}$ = 7 TeV using AMPT. The value of this ratio is explicitly higher for the $\alpha$-clustered structure in the most central case and fluctuates around the Woods-Saxon profile, consistent around unity in the mid-central to peripheral collisions. This demonstrates a balance between the geometry of the collisions and the fluctuations in the corresponding nucleon distributions. The exceptionally high value of $\langle\epsilon_3\rangle/\langle\epsilon_2\rangle$ in the most central collisions is limited to the $\alpha$-clustered structure. Therefore, for a hydrodynamical evolution, observation of unprecedented high value of $\langle v_3\rangle/\langle v_2\rangle$ in the most central O-O collisions could be a possible signature of such $\alpha$-clustered structure in the oxygen nuclei.\\

The observed high value of $\langle\epsilon_3\rangle/\langle\epsilon_2\rangle$ in the most central O-O collisions for $\alpha$-clustered structure compared to Woods-Saxon nuclear density profile can be understood by studying the distribution of $\epsilon_2$ and $\epsilon_3$ for both the nuclear profiles separately. Figure~\ref{eccdist} shows the eccentricity and triangularity distribution of most central (0-5)\% cases in O-O collisions at $\sqrt{s_{\rm NN}}$ = 7 TeV for both Woods-Saxon and $\alpha$-cluster density profiles estimated in AMPT. 
In Fig.~\ref{eccdist}, the eccentricity distribution is represented as the solid markers, and the triangularity distribution is represented as open markers. At the same time, the top and bottom panels of Fig.~\ref{eccdist} show the cases with the Woods-Saxon and $\alpha$-cluster density profiles, respectively. In the Woods-Saxon case, one observes that both the eccentricity and triangularity have their peaks shifted towards the lower values, indicating relatively isotropic distributions of the participants in the transverse plane. The eccentricity distribution for the $\alpha$-clustered structure has comparatively less mean and standard deviation compared to the Woods-Saxon case, showing a relatively more isotropic distribution of participants in the $\alpha$-clustered structure than the  Woods-Saxon case. On the other hand, the distribution of triangularity in the case of the $\alpha$-clustered structure is broader compared to the distribution of triangularity in the Woods-Saxon profile. It has a considerably higher mean value and standard deviation. This implies that even if the participant distribution in the $\alpha$-clustered structure is more isotropic in shape, it has more in-built fluctuations inside. These features of the interplay between eccentricity and triangularity with respect to different nucleon distribution profiles could be studied using different correlation functions, such as the normalized symmetric cumulants discussed in the following subsection.

\begin{figure}[ht!]
\includegraphics[scale=0.3]{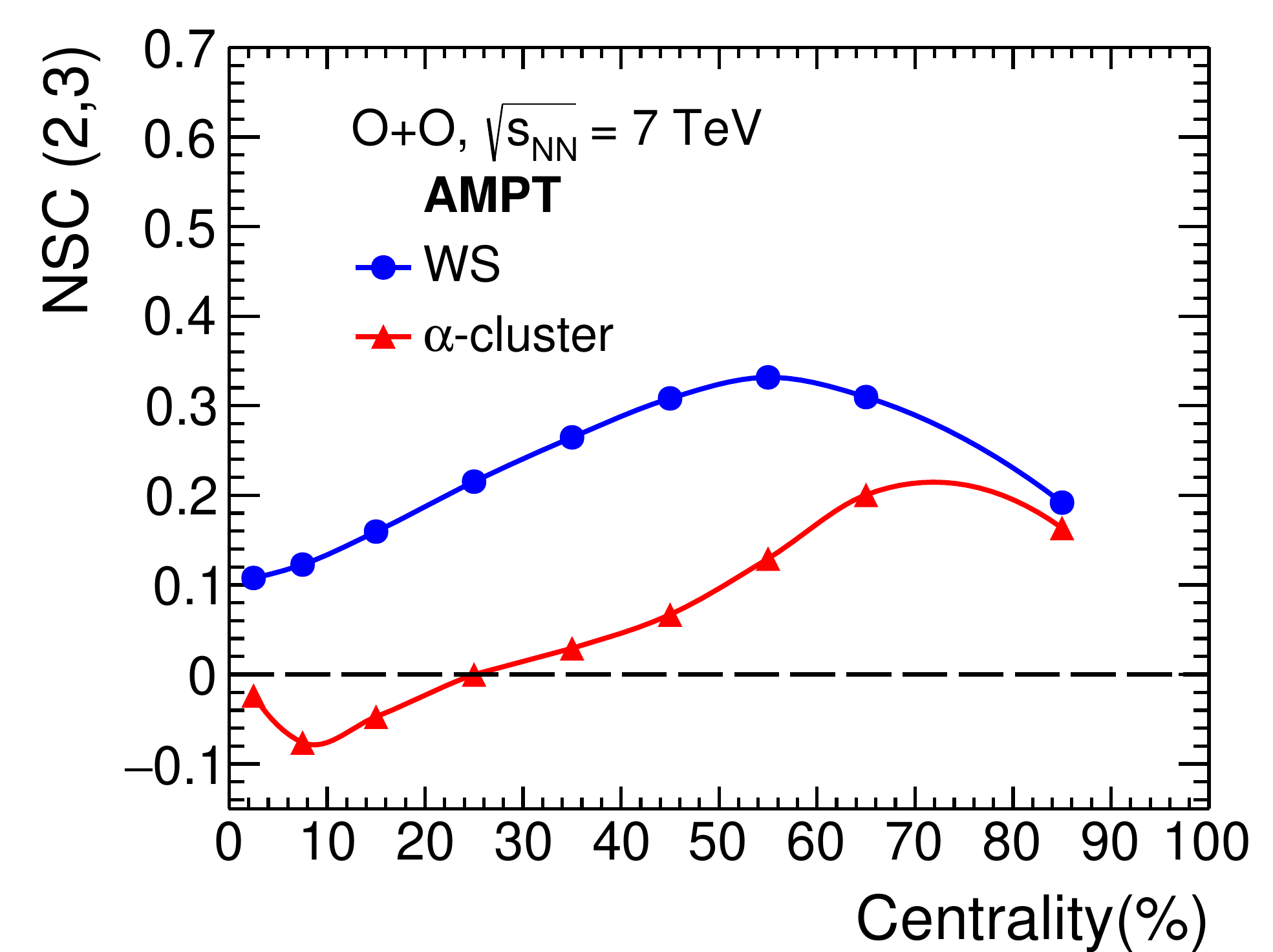}
\caption[]{(Color online) The normalized  symmetric cumulants coefficient NSC(2,3) as a function of centrality for both Woods-Saxon and $\alpha$-cluster type nuclear density profiles in O-O collisions at $\sqrt{s_{\rm NN}} = 7$~TeV in AMPT string melting model. }
\label{nsc}
\end{figure}

\begin{figure*}[ht!]
\includegraphics[scale=0.28]{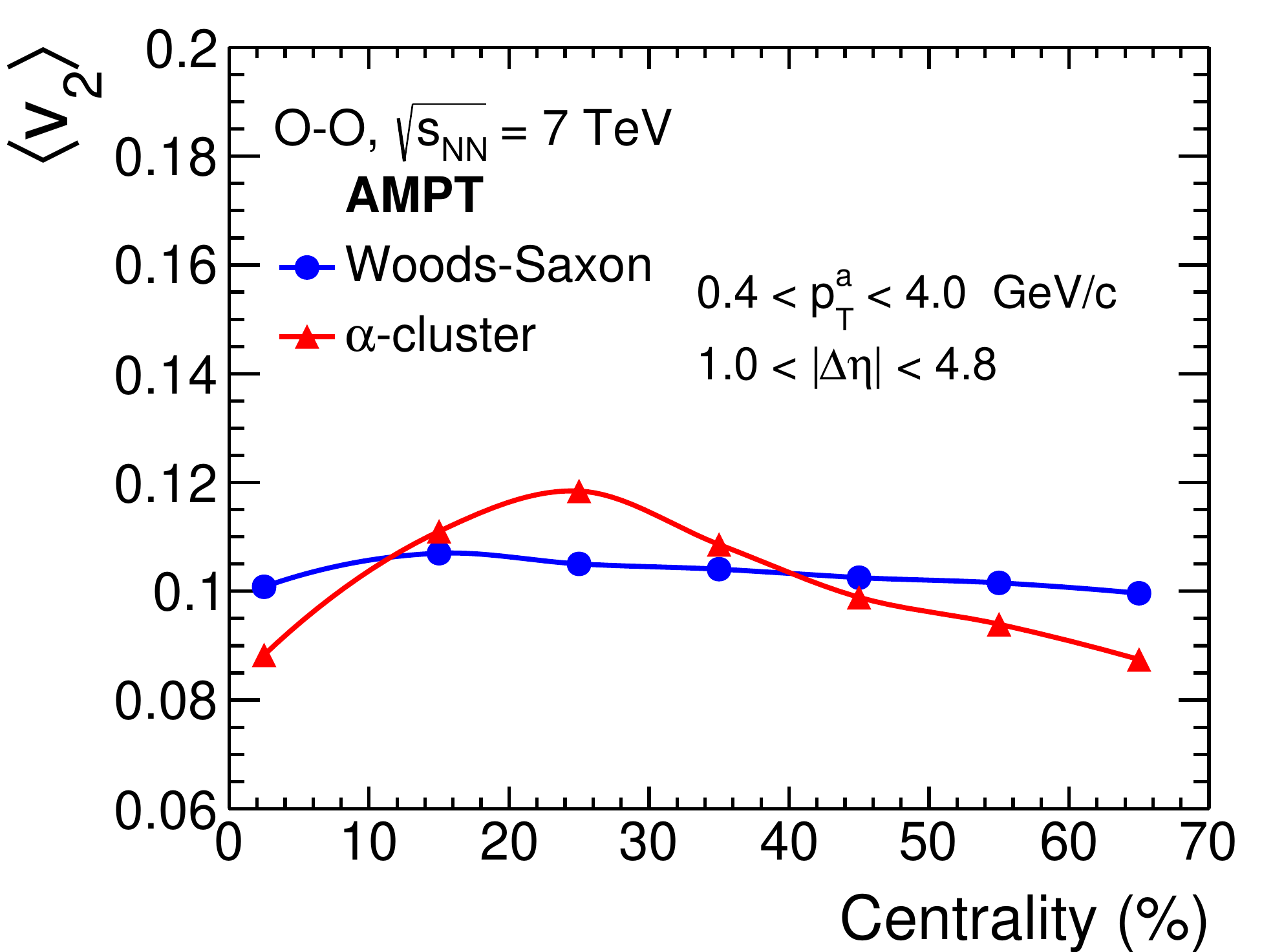}
\includegraphics[scale=0.28]{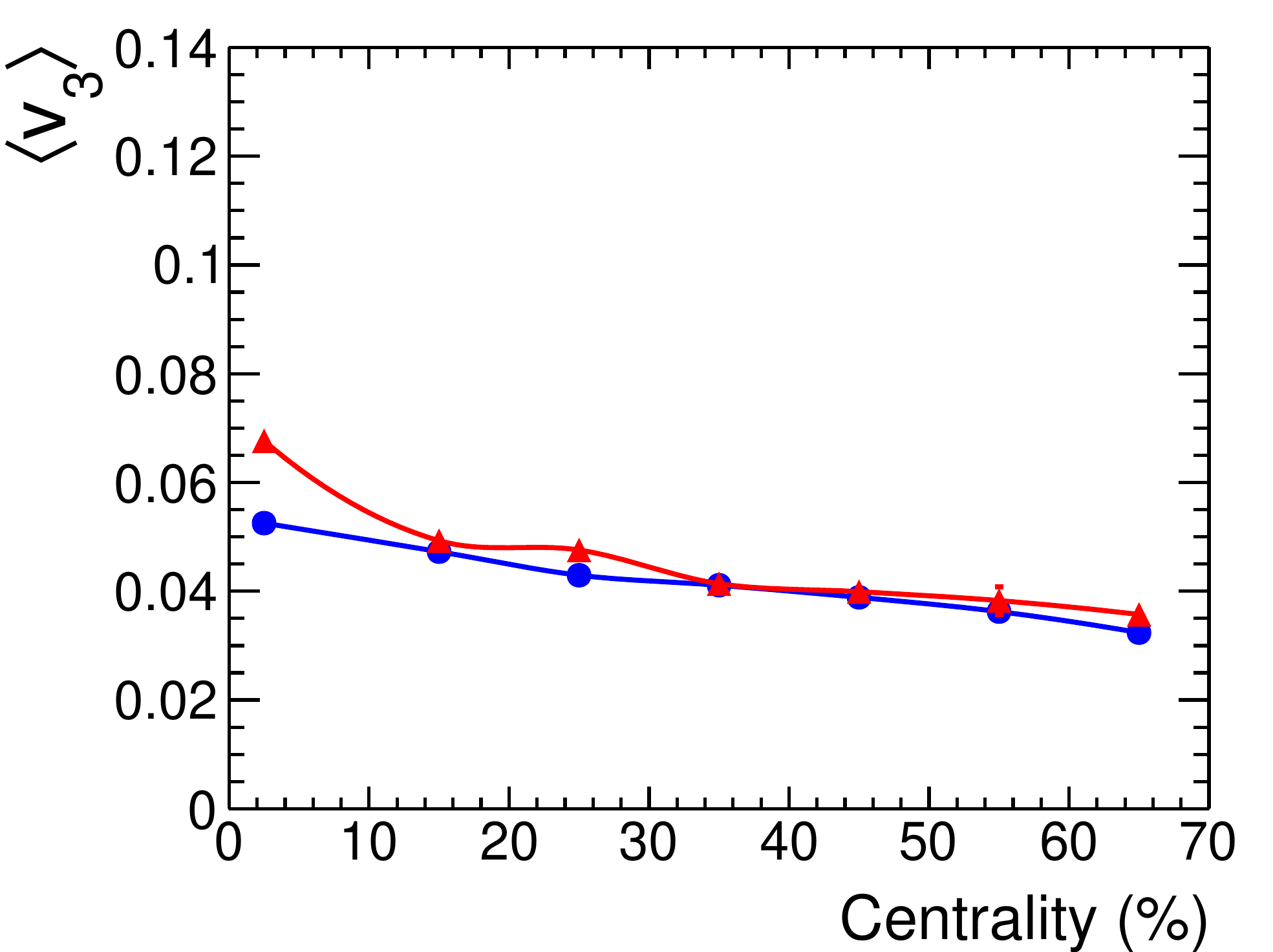}
\includegraphics[scale=0.28]{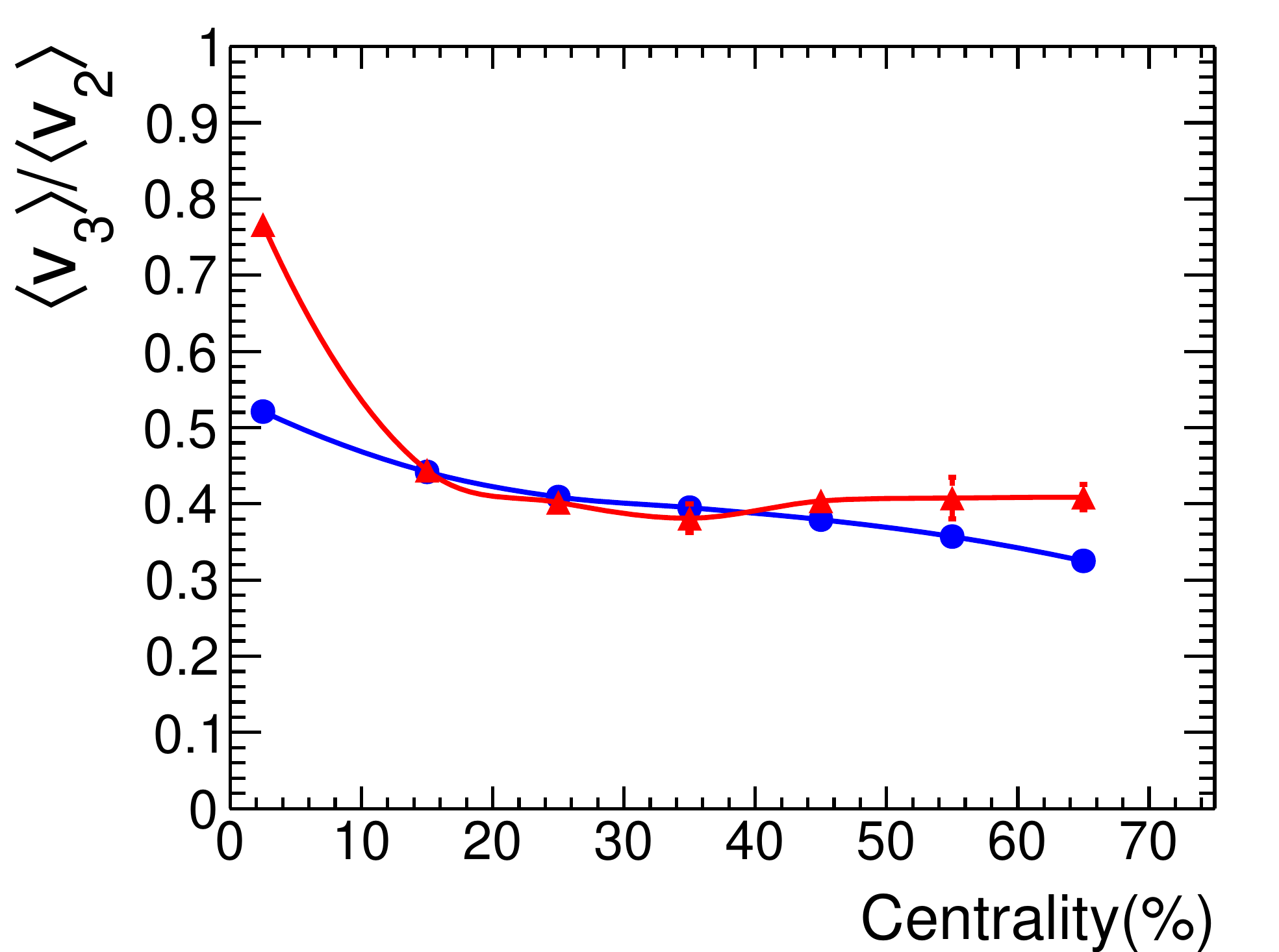}
\caption[]{(Color online) Integrated elliptic flow ($\langle v_{2}\rangle$) (left), triangular flow ($\langle v_{3}\rangle$) (middle), and the ratio  $(\langle v_{3}\rangle / \langle v_{2}\rangle )$ (right) as a function of centrality for both Woods-Saxon and $\alpha$-cluster type nuclear density profiles in O-O collisions at $\sqrt{s_{\rm NN}} = 7$~TeV from AMPT model. }
\label{v2}
\end{figure*}

\subsection{Normalized symmetric cumulants NSC(n,m)}

To quantify the positive or negative correlations between eccentricity and triangularity for different nuclear density profiles with respect to collision centrality, we define normalized symmetric cumulants coefficient NSC(n,m), which is given by \cite{Dasgupta:2020orj}:

\begin{eqnarray}
\rm NSC(n,m)=\frac{\langle{\varepsilon_n}^2{\varepsilon_m}^2\rangle-\langle{\varepsilon_n}^2\rangle \langle{\varepsilon_m}^2\rangle}{\langle\varepsilon_n^2\rangle\langle\varepsilon_m^2\rangle}.
\label{nsceq}
\end{eqnarray} 

Figure~\ref{nsc} shows the normalized symmetric cumulants coefficient as a function of centrality for both Woods-Saxon and $\alpha-$clustered cases in O-O collisions at $\sqrt{s_{\rm NN}} = 7$~TeV in AMPT string melting model. The negative coefficient values represent the anti-correlation between the two variables. We see a negative correlation for the $\alpha$-clustered case up to mid-central (20--30\%). This suggests that there is an anti-correlation between 
$\langle \epsilon_2 \rangle$ and $\langle \epsilon_3 \rangle$ only in the case of $\alpha$-clustering. However, we got positive correlations for the Woods-Saxon density profile for all centralities.  It should be noted that several recent studies investigate the properties of the initial-state eccentricities and final hadron flow observables from the collisions of clustered carbon and heavy ions at various beam energies in the event-by-event framework {\cite{Dasgupta:2020orj, Broniowski:2013dia}. 

\subsection{Elliptic flow and triangular flow}

Figure~\ref{v2} shows the $p_{\rm T}$-integrated elliptic (left) and triangular flow (middle) and the ratio of triangular to the elliptic flow coefficient (right) as a function of collision centrality in O-O collisions at $\sqrt{s_{\rm NN}}$ = 7 TeV for both Woods-Saxon density profile and $\alpha$-clustered structure using AMPT. The elliptic flow for the Woods-Saxon density profile does not have a strong centrality dependence, whereas the $\alpha$-clustered structure is observed to have significant centrality dependence. Unlike the Woods-Saxon profile, where the elliptic flow is finite yet almost flat with centrality, in $\alpha$-clustered structure, the elliptic flow value increases as one moves initially from central to midcentral collisions, and then attains a maximum around (20-30)\% centrality class. Thereafter, the value decreases towards the peripheral collisions. On the other hand, triangular flow for both Woods-Saxon density profile and $\alpha$-clustered structure have similar trends, i.e., maximum at the central collisions and decreases towards the peripheral collisions. This structure of triangular flow is peculiar to observe, considering the heavy-ion-like behavior where the value for the triangular flow peaks at the mid-central collisions. The cause of the peculiar behavior of elliptic and triangular flow may be due to the fact that the smaller system size and shorter lifetime of the fireball do not fully help transform the initial eccentricities to the final-state anisotropic flow coefficients. It is to be noted that the $\alpha$-clustered structure has a more significant triangular flow compared to the Woods-Saxon density profile throughout the centrality classes. The observations reported in Ref~\cite{Li:2020vrg} also present higher values of triangular flow for $\alpha$-clustered structure as compared to the Woods-Saxon density profile. However, Ref.~\cite{Li:2020vrg} reports lesser values for both elliptic and triangular flow for both the nuclear density profiles as they employ a different set of kinematic cuts, i.e., $-0.5<y<0.5$ and $(0.2<p_{\rm T}<3)$ GeV/c. Furthermore, the studies based on the hydrodynamic simulations in Ref.~\cite{Ding:2023ibq} show a minimal dependence of the $\alpha$-clustered structure on the elliptic and triangular flow coefficients. In the right plot of Fig.~\ref{v2}, where $\langle v_{3}\rangle / \langle v_{2}\rangle$ as a function of centrality is shown, one observes the ratio to be below one throughout the centrality classes and for both the density profiles and as one goes towards the peripheral collisions, the ratio seems to decrease for both the nuclear profiles. The value of $\langle v_{3}\rangle / \langle v_{2}\rangle$ for $\alpha$-clustered structure is larger than the Woods-Saxon density profile towards the most central and peripheral cases. Interestingly, one observes a sharp hike in the $\langle v_{3}\rangle / \langle v_{2}\rangle$ value for the most central case, which is inferred from the right panel of Fig.~\ref{ecc}, i.e., $\langle\epsilon_3\rangle/\langle\epsilon_2\rangle$ vs. centrality. This might be a possible signature of $\alpha$-clustered structure of oxygen nuclei in O-O collisions which can be verified in future experimental studies.\\

%The elliptic flow almost retains the trend from the eccentricity for $\alpha$-clustered structure and Woods-Saxon profile for a given centrality; however, it does not preserve the centrality dependence from the eccentricity, as the number of participants is significantly less in the case of O-O collisions. As opposed to the behavior of triangularity between the two nuclear structures, triangular flow is observed to have opposite values, i.e., triangular flow values for $\alpha$-clustered structure are more significant than the Woods-Saxon density profile. 

\begin{figure}[ht!]
\includegraphics[scale=0.3]{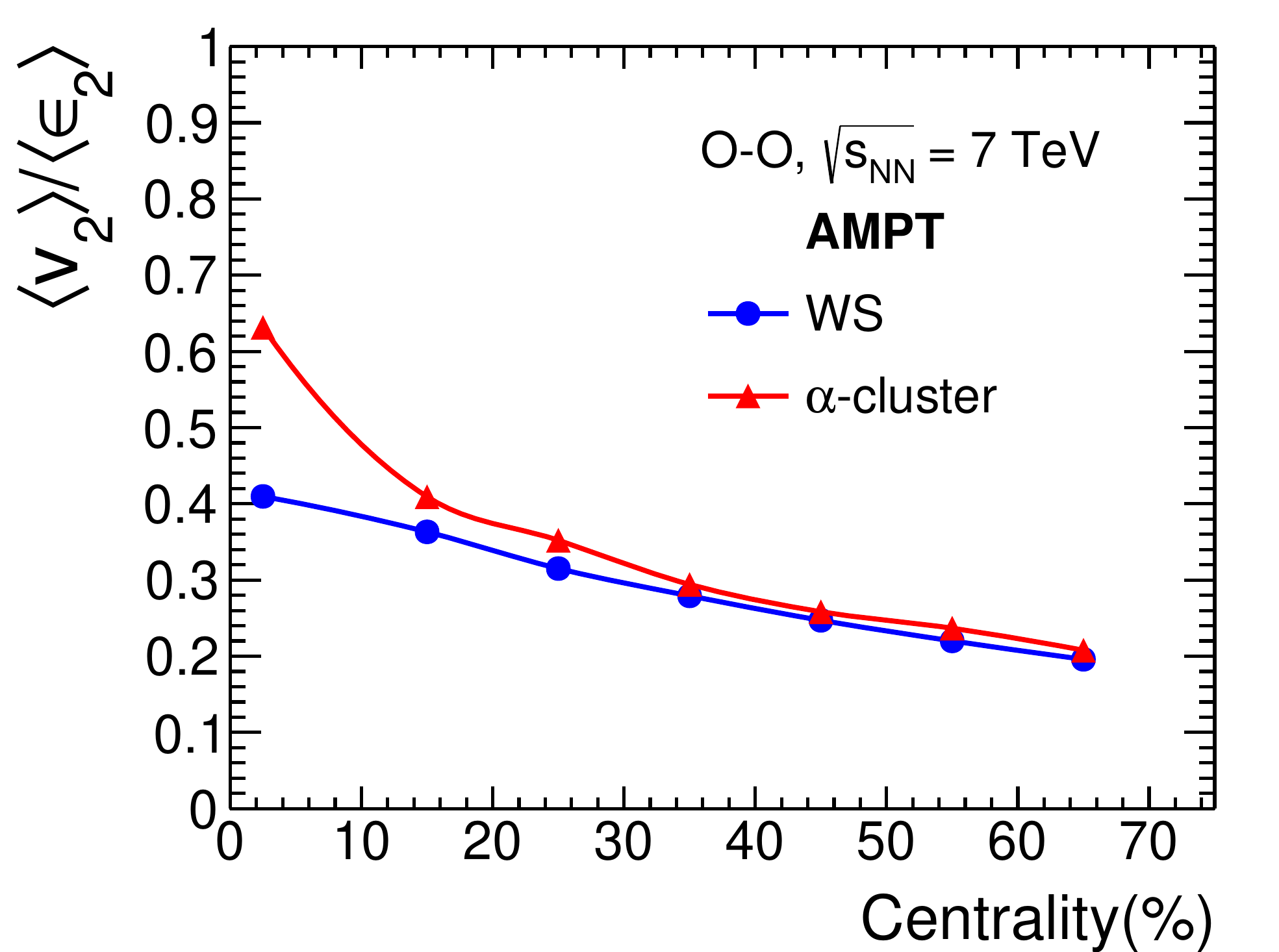}
\includegraphics[scale=0.3]{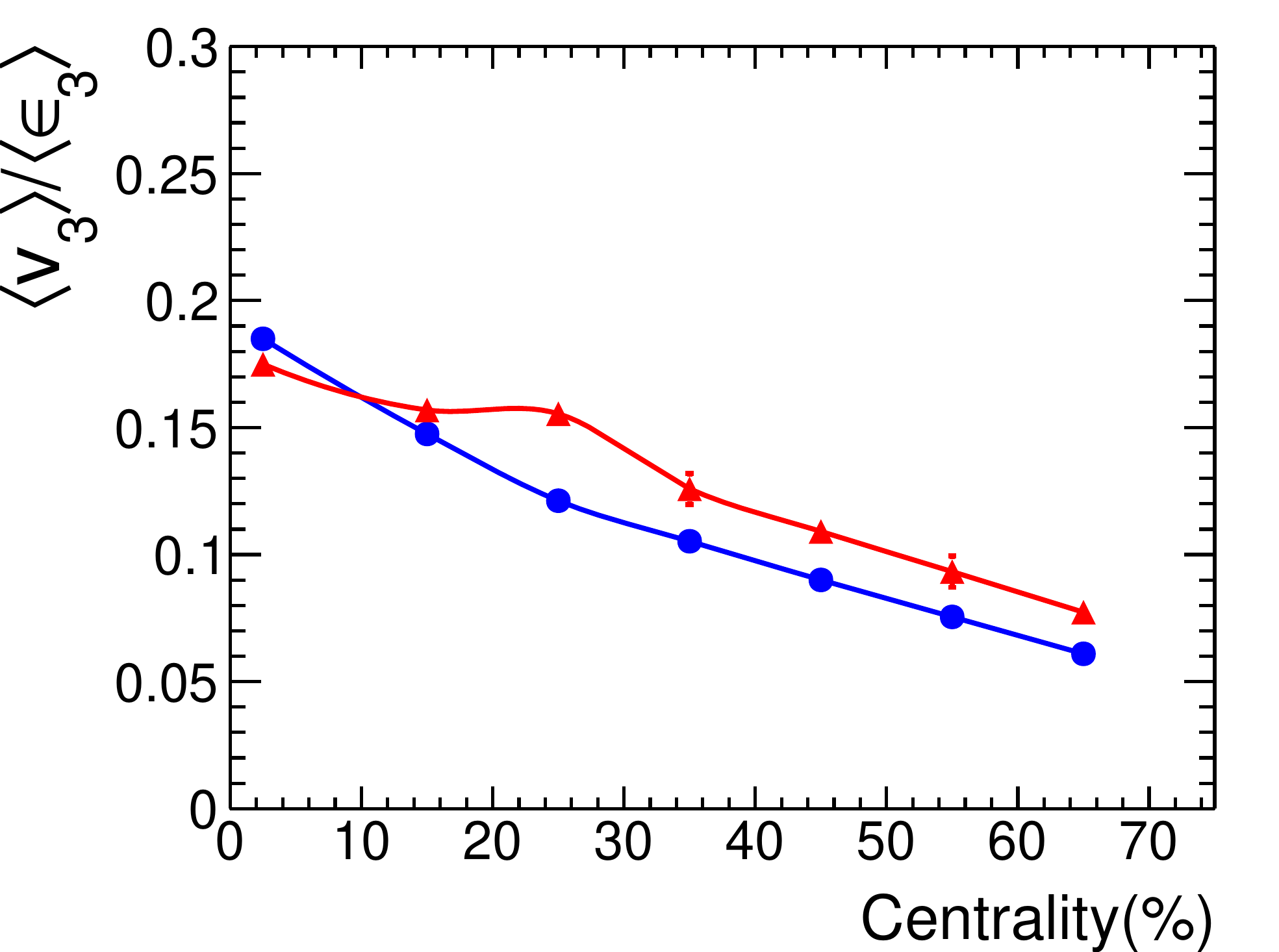}
\caption[]{(Color online)  The ratio $\langle v_{2}\rangle / \langle\epsilon_{2}\rangle$ (top) and $\langle v_{3}\rangle / \langle\epsilon_{3}\rangle$ (bottom) for O-O collisions at $\sqrt{s_{\rm NN}} = 7$~TeV for both Woods-Saxon and $\alpha$-cluster type nuclear density profiles in AMPT. }
\label{v2bye2}
\end{figure}

\begin{figure*}[ht!]
\centering
\includegraphics[scale=0.85]{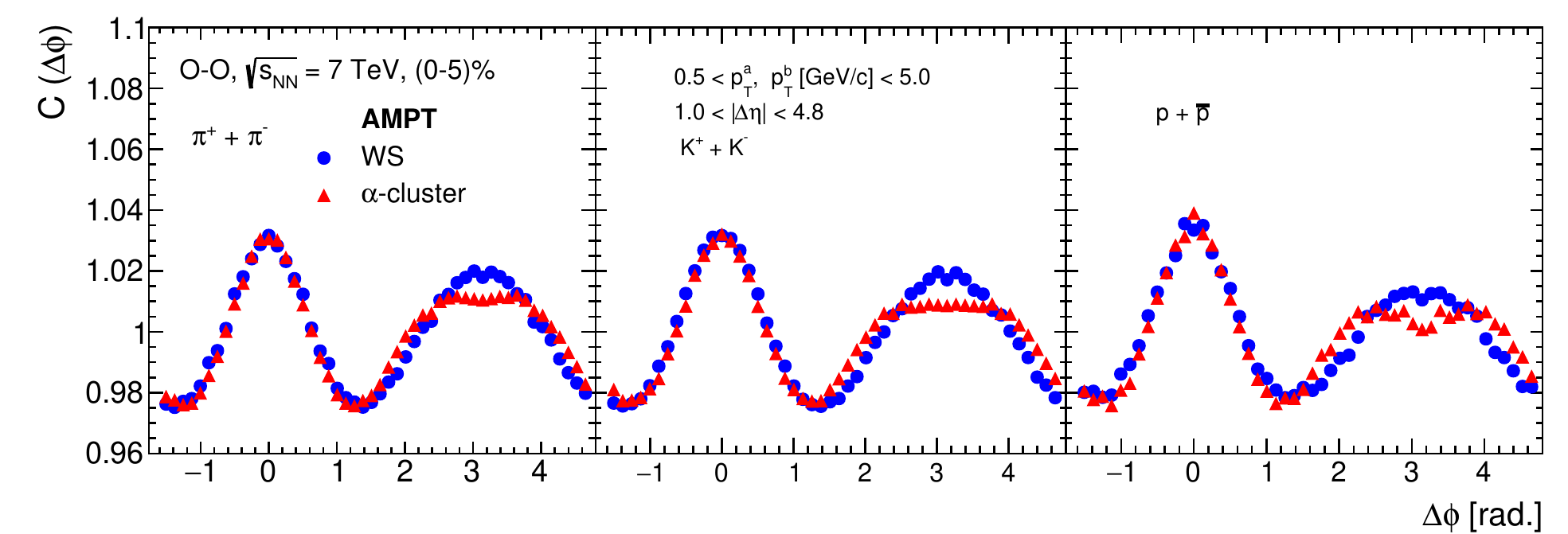}
\caption[]{(Color online) Two-particle azimuthal correlation function for $\pi^{\pm}$, $K^{\pm}$, and $p+\bar{p}$ in the most central O-O collisions at $\sqrt{s_{\rm NN}} = 7$~TeV using Woods-Saxon and $\alpha$-cluster type nuclear density profiles.}
\label{fig6}
\end{figure*}

\begin{figure*}[ht!]
\includegraphics[scale=0.28]{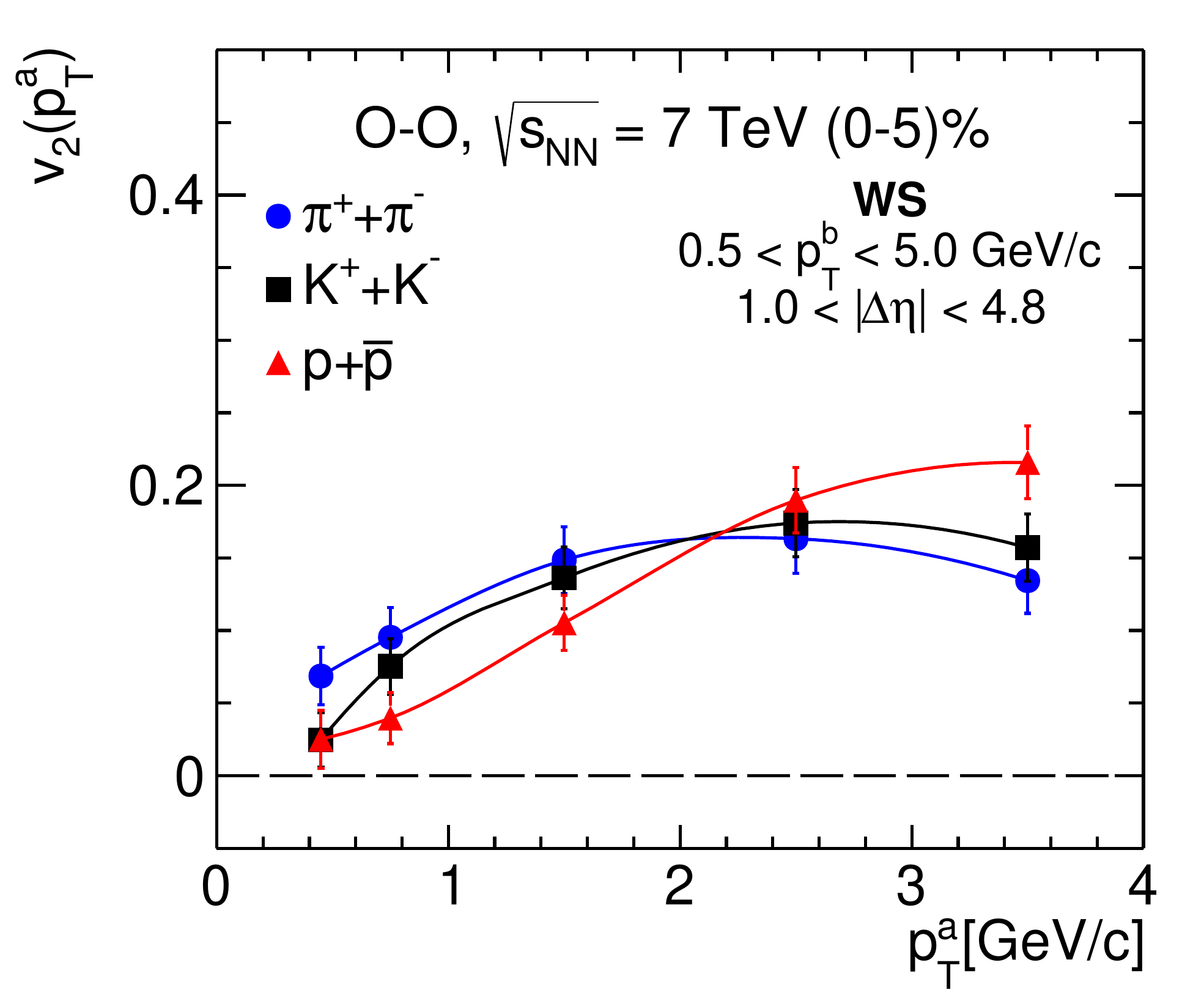}
\includegraphics[scale=0.28]{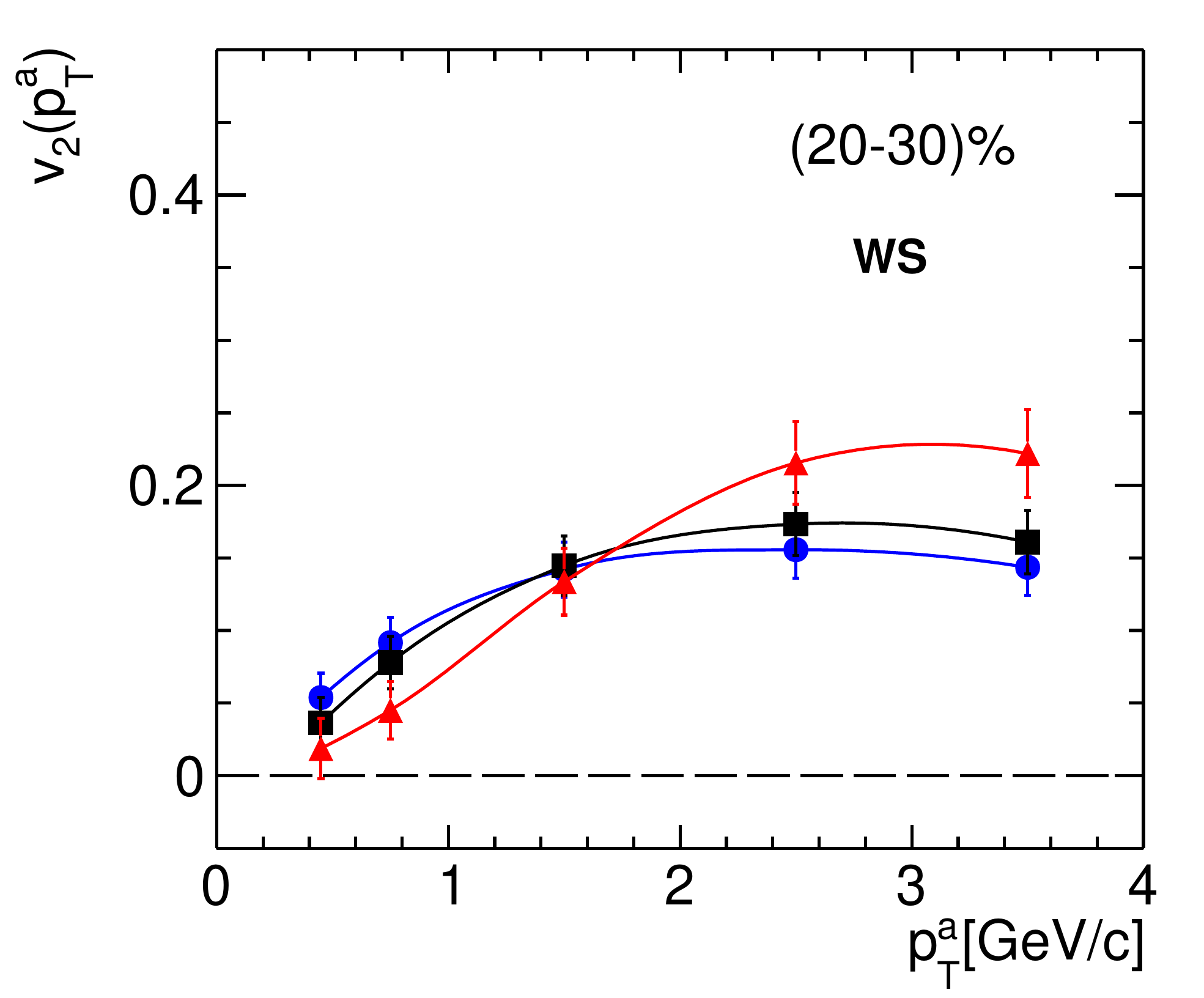}
\includegraphics[scale=0.28]{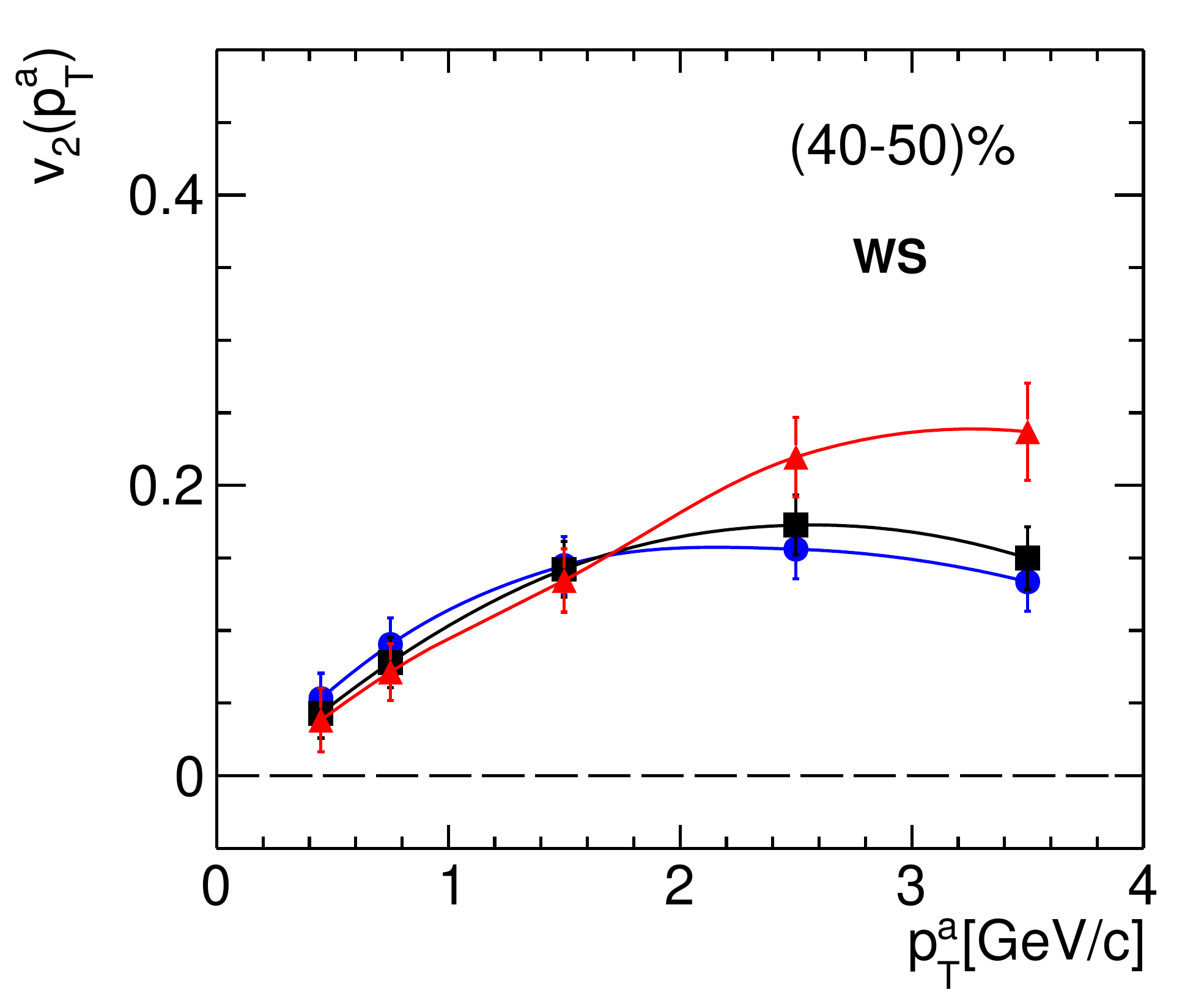}
\includegraphics[scale=0.28]{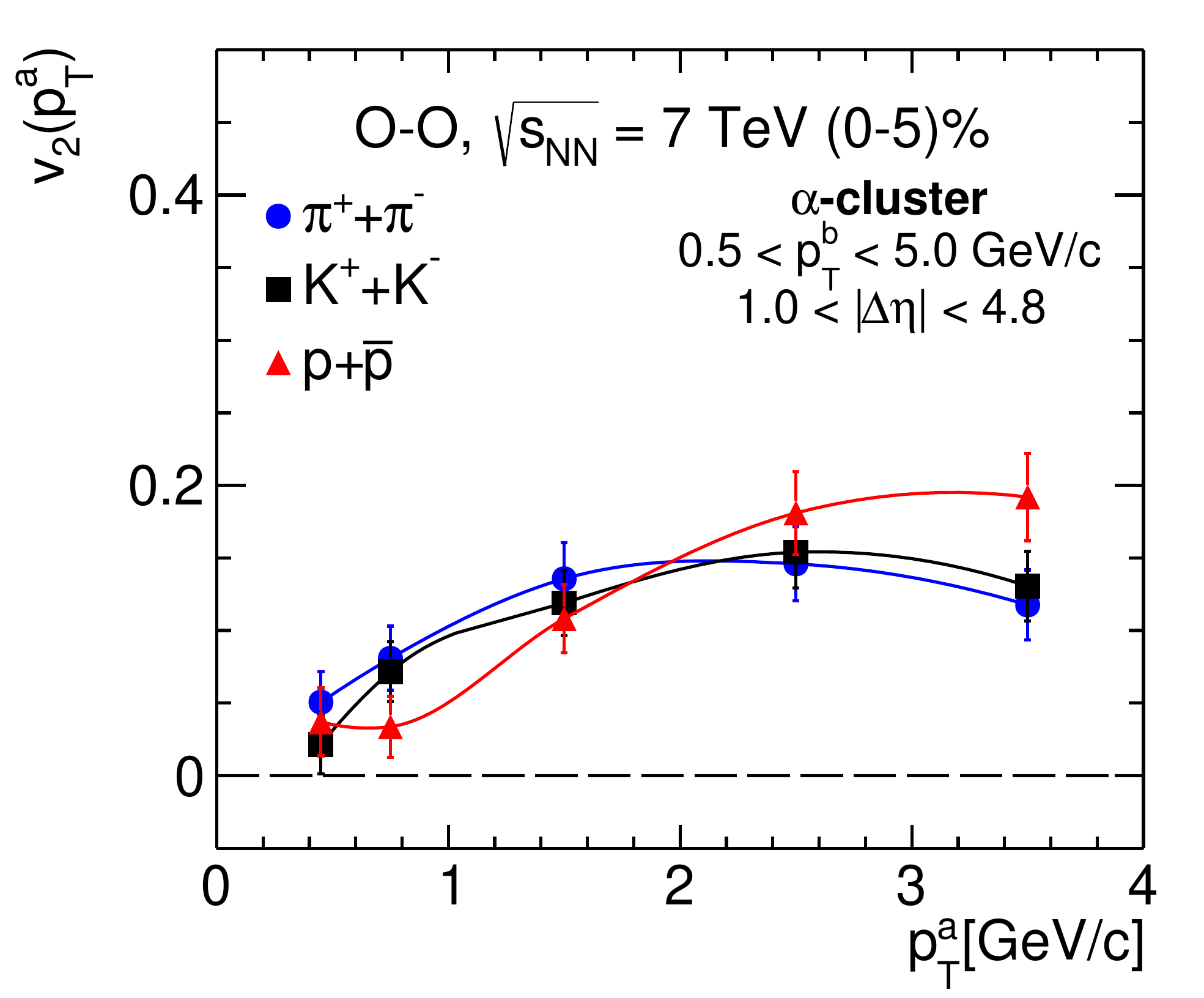}
\includegraphics[scale=0.28]{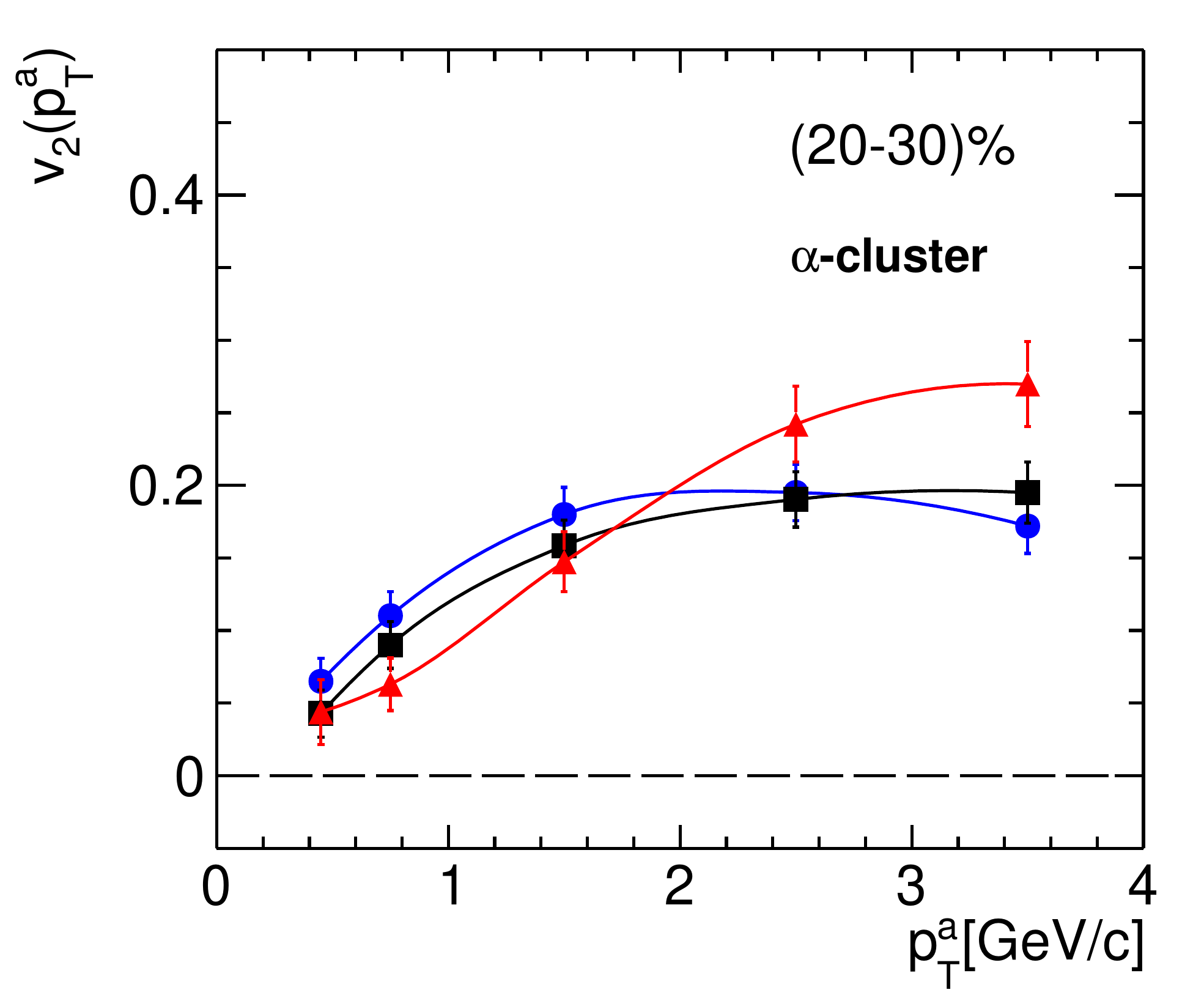}
\includegraphics[scale=0.28]{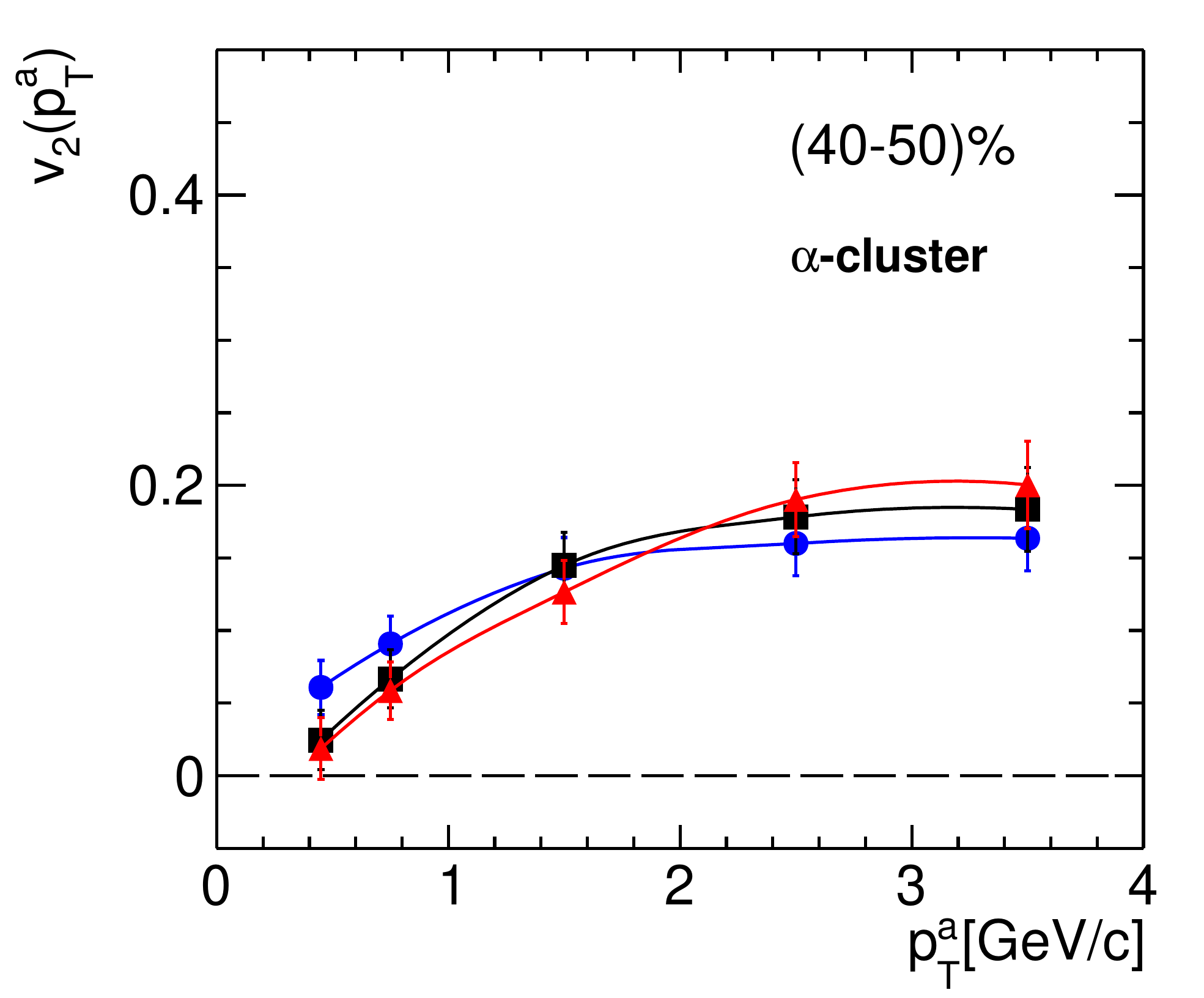}
\caption[]{(Color online) Transverse momentum $(p_{\rm T})$ dependence of $v_{2}(p_{\rm T})$ for $\pi^{\pm}$, $\rm K^{\pm}$, and $\rm p+\bar{p}$ in O-O collisions at $\sqrt{s_{\rm NN}} = 7$~TeV. Results include Woods-Saxon and $\alpha$-cluster type nuclear charge density profiles for the oxygen nucleus.}
\label{fig7}
\end{figure*}

In Fig.~\ref{v2bye2}, $\langle v_{2}\rangle / \langle \epsilon_{2}\rangle$ (top) and $\langle v_{3}\rangle / \langle \epsilon_{3}\rangle$ (bottom) as a function of centrality for $\alpha$-clustered structure and Woods-Saxon density profile for O-O collisions at $\sqrt{s_{\rm NN}}$ = 7 TeV are shown. In Fig.~\ref{v2bye2}, $\langle v_{2}\rangle / \langle \epsilon_{2}\rangle$ and $\langle v_{3}\rangle / \langle \epsilon_{3}\rangle$ is observed to be decreasing towards the peripheral collisions for both the nuclear profiles; however, both these ratios for the $\alpha$-clustered structure are larger as compared to the Woods-Saxon density profile. Both $\langle v_{2}\rangle / \langle \epsilon_{2}\rangle$ and $\langle v_{3}\rangle / \langle \epsilon_{3}\rangle$ tell about the effect of the medium on the evolution of the flow coefficients, i.e., $\langle v_{2}\rangle$ and $\langle v_{3}\rangle$ from initial eccentricities, i.e., $\langle \epsilon_{2}\rangle$ and $\langle \epsilon_{3}\rangle$, respectively. As discussed by the authors in Ref.~\cite{Schenke:2011bn}, it is known that anisotropic flow coefficients of different order are affected differently by the medium formed, and as the order of the flow coefficients increase, their sensitivity to the viscosity of the medium increase. Thus the observed enhanced values of $\langle v_{2}\rangle / \langle \epsilon_{2}\rangle$ and $\langle v_{3}\rangle / \langle \epsilon_{3}\rangle$ for $\alpha$-clustered structure compared to Woods-Saxon density profile may be attributed to a longer duration of the partonic or hadronic phase of the collision system.

\begin{figure*}[ht!]
\includegraphics[scale=0.28]{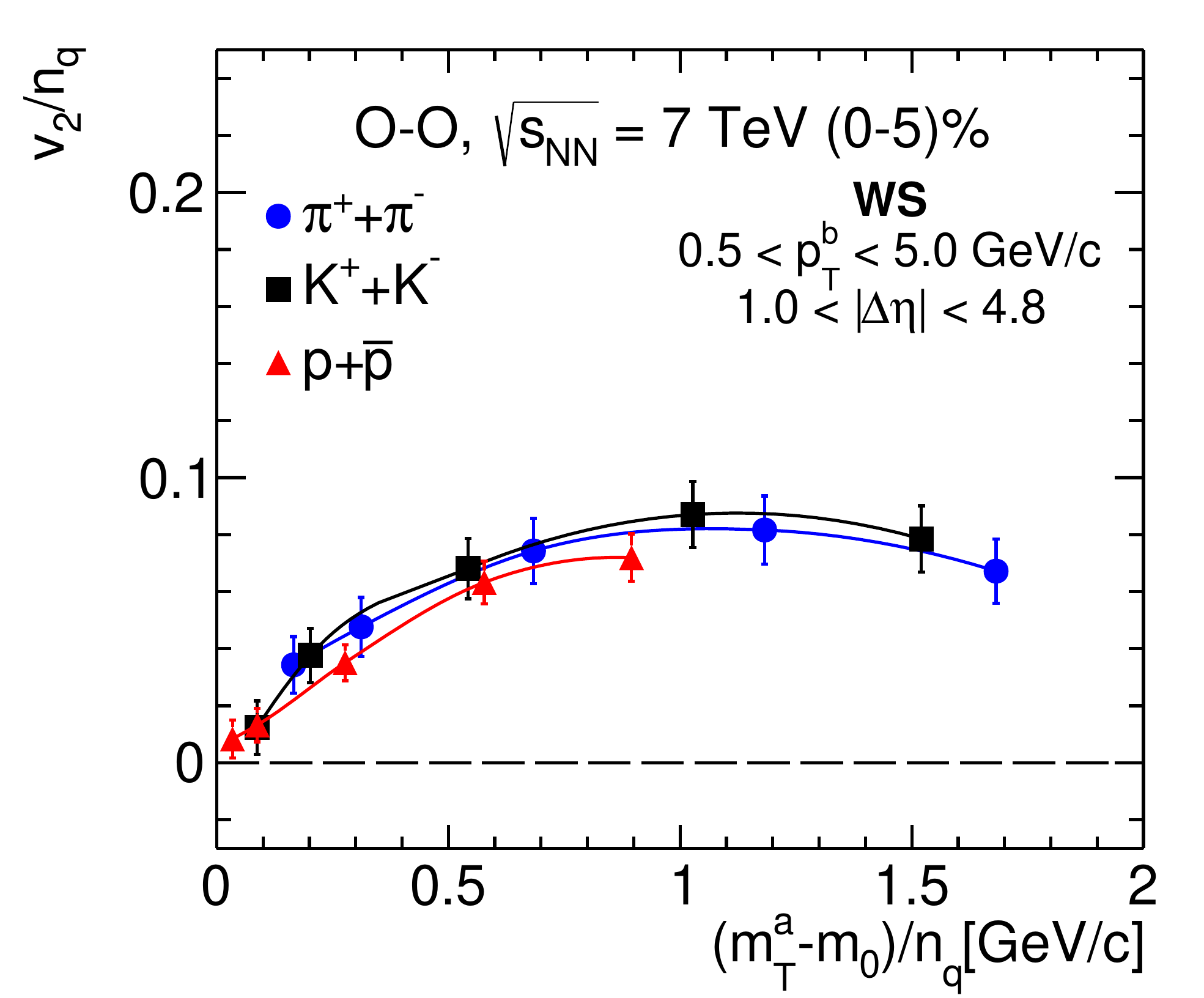}
\includegraphics[scale=0.28]{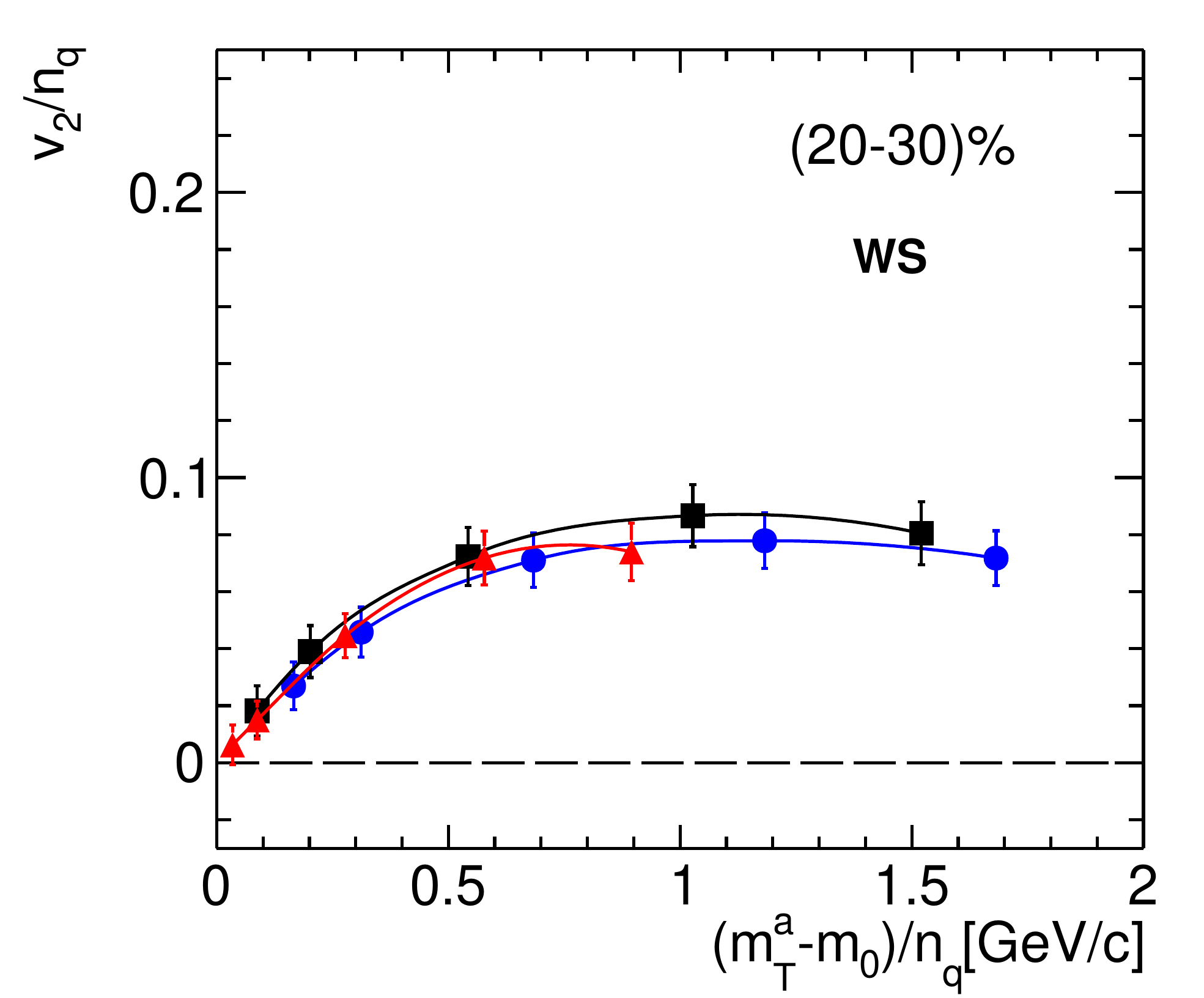}
\includegraphics[scale=0.28]{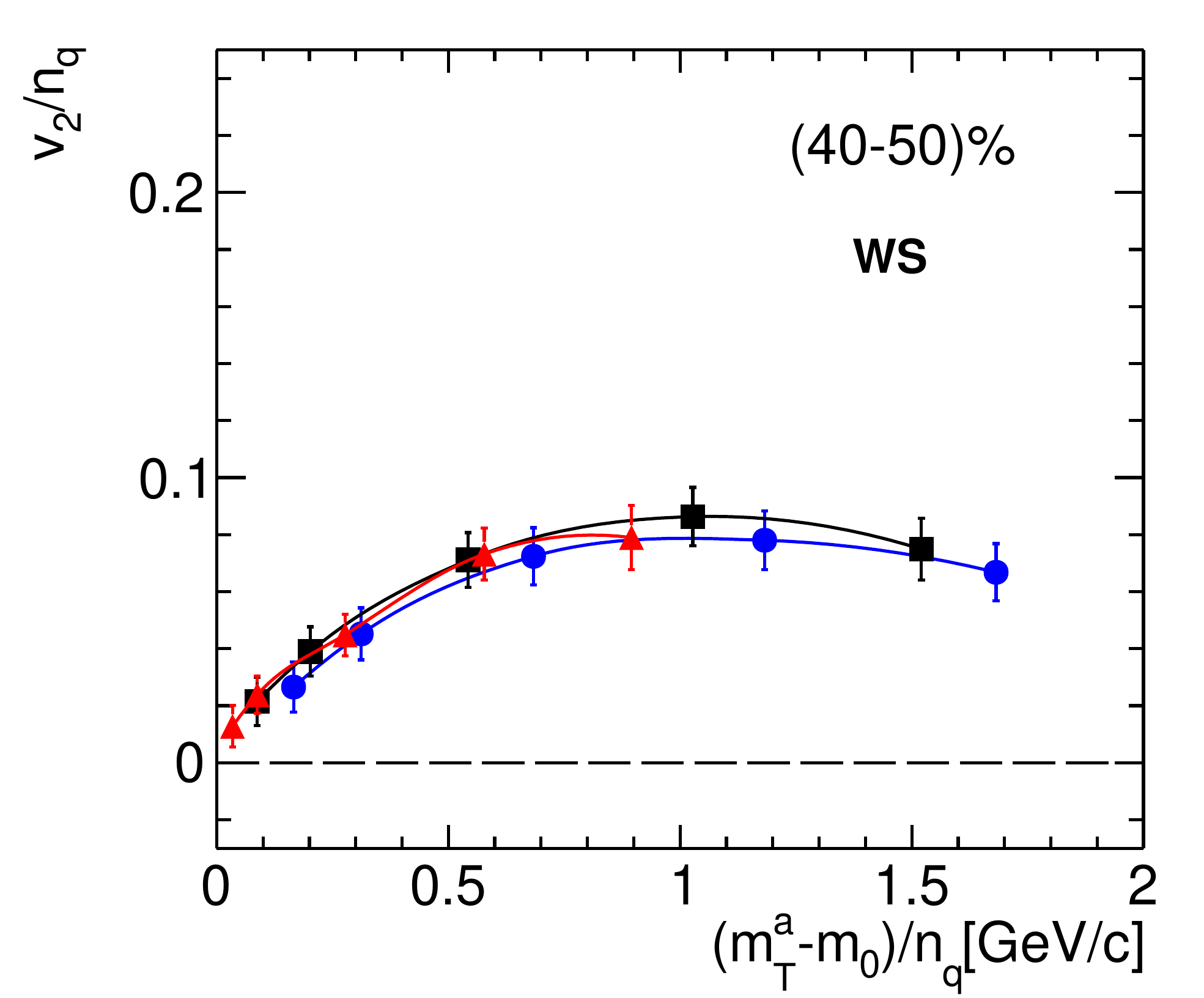}
\includegraphics[scale=0.28]{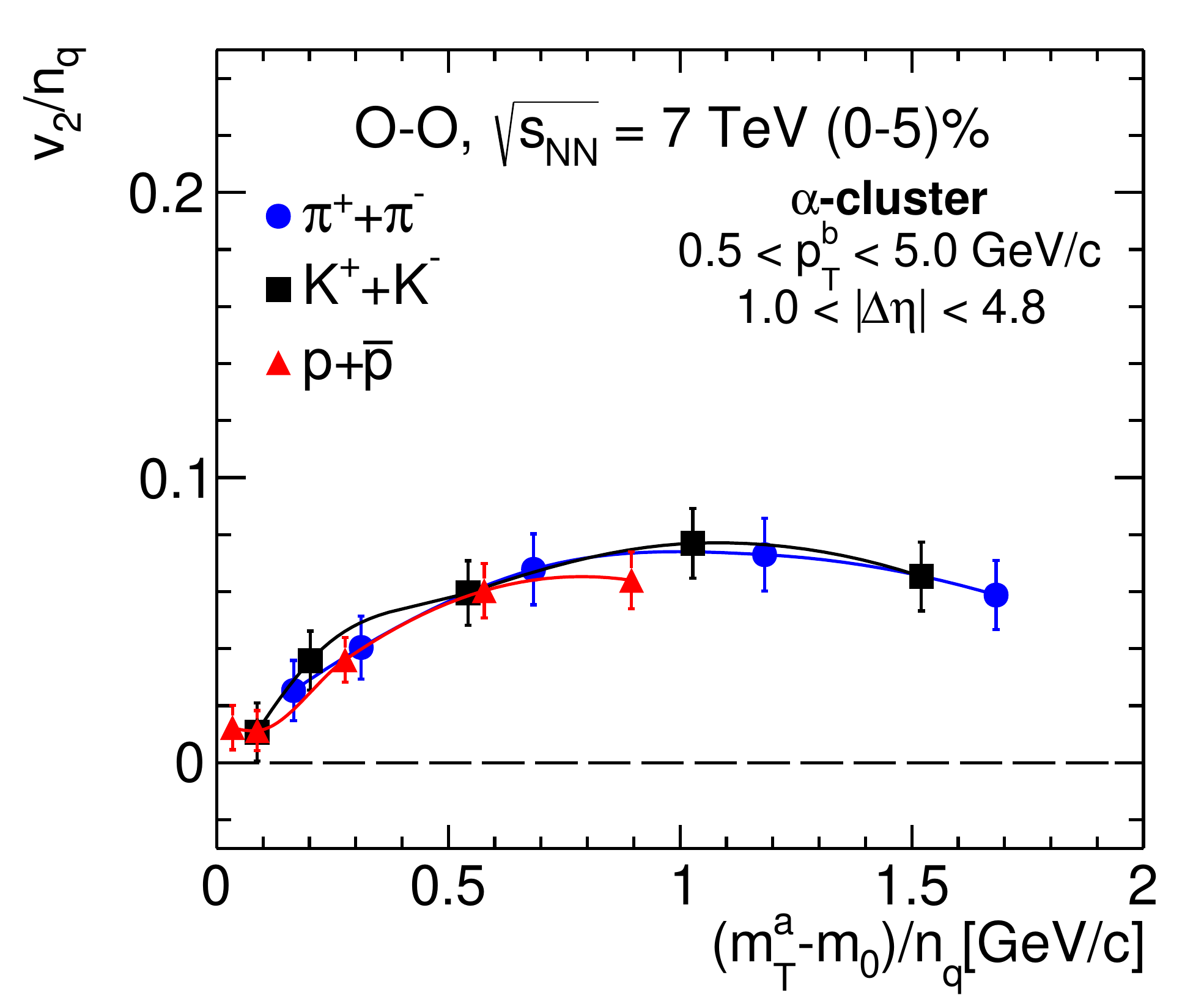}
\includegraphics[scale=0.28]{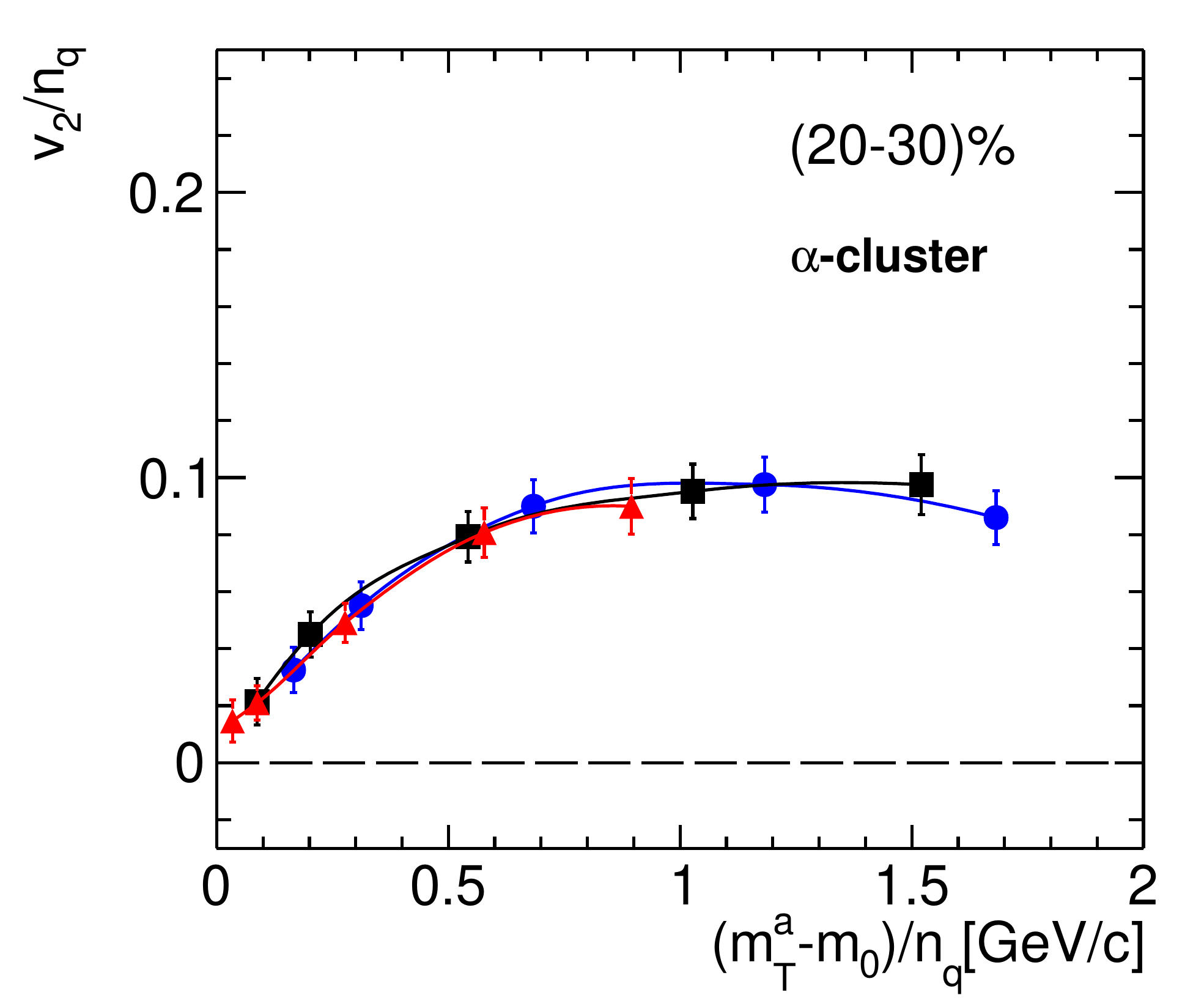}
\includegraphics[scale=0.28]{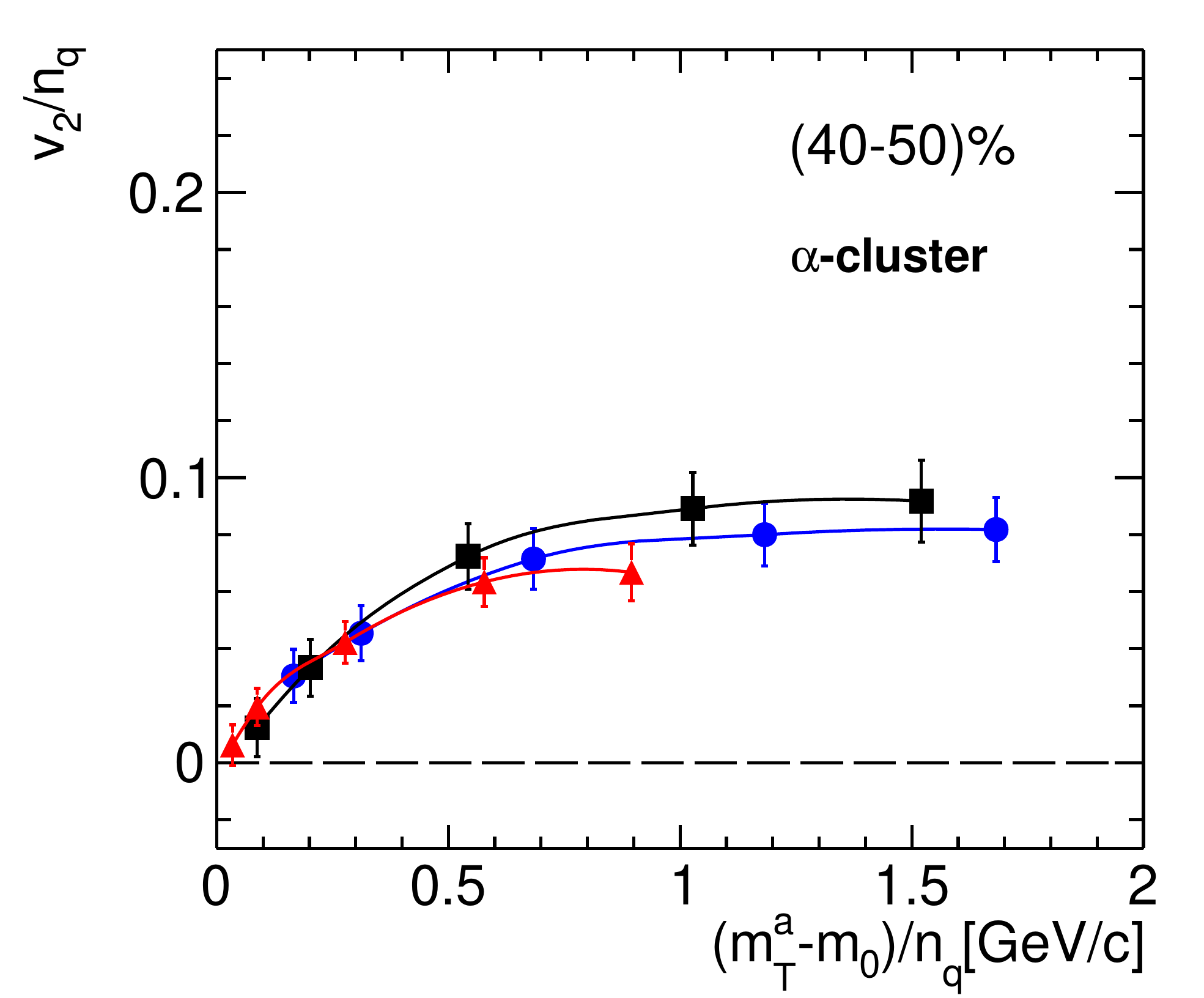}
\caption[]{(Color online) Transverse kinetic energy dependence of constituent-quark scaling of $v_2$ for $\pi^{\pm}$, $\rm K^{\pm}$, and $\rm p+\bar{p}$ in O-O collisions at $\sqrt{s_{\rm NN}} = 7$~TeV.}
\label{fig8}
\end{figure*}

\subsection{Elliptic flow of light-flavor hadrons and NCQ scaling}

Figure~\ref{fig6} shows the two-particle azimuthal correlation function ($C(\Delta\phi)$) for $\pi^{\pm}$, $K^{\pm}$, and $p+\bar{p}$ in the most central O-O collisions at $\sqrt{s_{\rm NN}} = 7$~TeV in the relative azimuthal angle $\Delta\phi \in [-\pi/2,3\pi/2]$. The blue dots and the red triangles represent the cases for the nucleus having the Woods-Saxon and tetrahedral $\alpha$-cluster density profiles respectively. The correlation function is constructed in the transverse momentum range, $0.5<p_{\rm T}^{\rm a},p_{\rm T}^{\rm b}<5.0$~GeV/{\it c}, in the relative pseudorapidity gap $1.0<|\Delta\eta|<4.8$. This pseudorapidity gap ensures the removal of short-range resonance decays and mini-jets contributing to the non-flow effects. The magnitude of the peak of the correlation function is related to the magnitude of the anisotropic flow coefficients. Both the density profiles show similar magnitudes of peaks at the near side ($\Delta\phi \simeq 0$). However, for the case with $\alpha$-cluster, there is an away-side ($\Delta\phi \simeq \pi$) broadening and suppression in the two-particle azimuthal correlation function. This effect gets more pronounced as one moves from pion to kaon and then to proton. This away-side valley may arise due to the more violent interactions among the partons caused due to the more compact and denser fireball created in nuclear collisions having $\alpha$-clusters, which also results in higher multiplicity than the Woods-Saxon case in similar centrality bins~\cite{Behera:2021zhi}. The presence of two peaks on the away side adds to this understanding as it leads to an enhanced contribution to the triangular flow~\cite{Mallick:2021hcs, Prasad:2022zbr}. In short, by comparing the $C(\Delta\phi)$ distributions of an ordinary Woods-Saxon nucleus with the $\alpha$-clustered nucleus, one can observe a dependence of the azimuthal correlation function on the initial density profile of the nucleus. These results are in line with the observations reported in Ref.~\cite{Wang:2021ghq}. It can be noted that there might be residual jetlike correlations leading to the away-side signal suppression. Proton being the massive one, shows a relatively higher suppression in the medium than kaon and pion. \\

Figure~\ref{fig7} shows centrality dependence of single particle elliptic flow coefficients for $\pi^{\pm}$, $K^{\pm}$, and $p+\bar{p}$ in O-O collisions at $\sqrt{s_{\rm NN}} = 7$~TeV, for Woods-Saxon and $\alpha$-clustered density profiles. Three centrality bins are chosen for this study, the most central (0--5)\%, intermediate (20--30)\%, and noncentral (40--50)\%. For the Woods-Saxon case, there is a very weak dependency of $v_2(p_{\rm T})$ on centrality for the three particle types. But for the $\alpha$-clustered case, in (20--30)\% centrality, there is a higher  $v_2(p_{\rm T})$ as compared to the other centrality bins. In the Woods-Saxon case, we argue that the smaller system size does not allow much variation in $v_{2}$ as a function of centrality, irrespective of an increasing $\epsilon_{2}$;  however, for the $\alpha$-cluster case, the more compact geometry tends to produce comparatively a denser medium, and thus the variation of $v_{2}$ with respect to centrality comes into picture. Now moving onto the particle types, at low $p_{\rm  T}$, there is a distinct mass ordering in the elliptic flow of $\pi^{\pm}$, $K^{\pm}$, and $p+\bar{p}$. This is understood to have originated from the competing effects of radial (symmetric) flow and anisotropic flow. In the intermediate $p_{\rm T}$, the baryon-meson flow separation occurs, with baryon $v_2$ being greater than that of the meson. This comes into existence due to the quark coalescence mechanism of hadronization embedded in the AMPT string melting model.\\

Figure~\ref{fig8} shows the centrality dependence of $v_{2}(p_{\rm T}^a$)/$n_q$ scaling as a function of $(m_{\rm T} - m_0)/n_q$ for $\pi^{\pm}$, $K^{\pm}$ and $p+\bar{p}$ in O-O collisions at $\sqrt{s_{\rm NN}} = 7$~TeV for both Woods-Saxon and $\alpha$-cluster type nuclear density profiles. Here, $n_q = 2$ for mesons, $n_q = 3$ for baryons, and the transverse kinetic energy, $KE_{\rm T} = (m_{\rm T} - m_0)$, where $m_{\rm T}$ is the transverse mass and $m_0$ is the rest mass of the particle. These plots quantitatively show the elliptic flow of the constituent quarks as a function of their transverse kinetic energy. As discussed earlier, within the AMPT framework, in Pb-Pb collisions at the LHC energies, the NCQ scaling is violated. However, at the same energy in Si-Si collision system, the NCQ scaling is found to be valid. In O-O collisions, which is an even smaller system, the scaling is valid for all centrality classes irrespective of the Woods-Saxon or $\alpha$-clustered type nucleus. Thus, the presence of $\alpha$-cluster geometry does not seem to play a role in the NCQ-scaling behaviour. However, the away-side broadening seen in Fig.~\ref{fig6}, seems to have been influenced from the early stages of the collisions due to the change in the nuclear density profiles leading to a more dense and compact system formation in the presence of $\alpha$-cluster type nucleus.

\section{Summary}
\label{section4}

In this paper, we have investigated the effect of Woods-Saxon and $\alpha$-clustered nuclear geometry on the eccentricity and triangularity along with their correlations, elliptic flow, triangularity flow, and NCQ scaling in O-O collisions at  $\sqrt{s_{\rm NN}} = 7$~TeV in the framework of a multiphase transport model. The key findings are  summarized below:

\begin{enumerate}[(i)]

\item Eccentricity and triangularity are found to vary with a change in the density profiles. However, the effects are more pronounced in the most central case, where the initial state has more triangularity than eccentricity for an $\alpha$-clustered oxygen nucleus as compared to the normal Woods-Saxon type distribution.

\item Employing the normalized symmetric cumulants, we observe that the strength of the correlation between eccentricity and triangularity for the Woods-Saxon density profile is more than for the $\alpha$-clustered structure. Also, the appearance of negative NSC (2,3) value for the $\alpha-$clustered nucleus in the most central cases is observed. 

\item In the Woods-Saxon type nucleus, the elliptic flow is found to depend weakly on the centrality of the collision. However, in the $\alpha-$clustered nucleus, the elliptic flow increases from central to mid-central collisions and then decreases while moving from mid to peripheral collisions.
 
\item We report an enhancement in the $\langle v_{3}\rangle / \langle v_{2}\rangle $ towards the most central collisions for the $\alpha-$clustered nucleus than the Woods-Saxon case.

\item The two-particle azimuthal correlation function $[C(\Delta\phi)]$ of the identified particles shows an away-side broadening for the $\alpha$-clustered type nucleus. This hints towards a denser and more compact system formation in the $\alpha$-clustered nucleus.

\item The NCQ scaling is valid for all centrality classes for both Woods-Saxon and $\alpha-$clustered type of nucleus. This observation is crucial as it hints towards the existence of a deconfined partonic medium in O-O collisions at $\sqrt{s_{\rm NN}} = 7$~TeV and the appearance of partonic collectivity.

\end{enumerate}

It would be interesting to compare these findings to experimental observations when experimental data are available in order to determine the density profile of the oxygen nucleus that is best suited to describe ultra-relativistic nuclear collisions. Although probing the nuclear density profile is a matter of low-energy nuclear scattering experiments, some of the 
observables in TeV nuclear collisions may be sensitive to the nuclear density profiles. In this study, we report a few 
such observables in heavy-ion collisions which could be sensitive to the nuclear density profiles and should be studied in
experimental data.

%Anisotropic flow ($v_{2}$) increased in  the mid-central region in the case of $\alpha$-clusterd density profile for all charged particles.

\section*{Acknowledgements}

D.B. acknowledges the financial support from CSIR, the Government of India. S.P. acknowledges the financial support from UGC,
the Government of India. R. S. sincerely acknowledges the DAE-DST, Government of India funding under the Mega-Science Project – “Indian participation in the ALICE experiment at CERN” bearing Project No. SR/MF/PS-02/2021-IITI (E-37123). The authors gratefully acknowledge the usage of resources of the LHC grid Tier-3 computing facility at IIT Indore.

\end{document}